\pdfoutput=1
\documentclass[a4paper,11pt]{article}
\usepackage{jheppub}
\usepackage[utf8]{inputenc}
\usepackage[T1]{fontenc}
\usepackage{amsmath,amssymb,bm}
\usepackage{booktabs,array}

\newcommand{\coderepo}{\url{https://github.com/migaelstrydom/holographic-vortex-critical-coupling}}
\newcommand{\AdS}{\mathrm{AdS}}
\newcommand{\astar}{\alpha_\star}

\title{A critical coupling for the magnetic instability of holographic $SU(2)$ plasmas}

\author{Migael Strydom}
\hypersetup{pdftitle={A critical coupling for the magnetic instability of holographic SU(2) plasmas},pdfauthor={Migael Strydom}}
\affiliation{Independent researcher, Cambridge, United Kingdom}
\emailAdd{migael@strydom.me.uk}

\abstract{We show that gravity can remove the linear magnetic instability of a holographic
plasma with an $SU(2)$ global symmetry. In the probe limit, charged vector
bosons become tachyonic in their lowest Landau level above a critical field
and condense into a vortex lattice. In Einstein--$SU(2)$ Yang--Mills theory
in $\AdS_5$ we include the backreaction of the field exactly, expanding only
in the condensate around the D'Hoker--Kraus magnetic brane. The critical
field diverges as the coupling $\alpha = \kappa^2/\hat g^2$ approaches
$\astar = 2/3$, above which the normal phase has no linear charged
instability at any field or temperature: we show this at zero temperature
and establish it numerically at finite temperature. At zero temperature the
field cancels from the mass of the lowest Landau level in the brane's
$\mathrm{AdS}_3 \times \mathbb{R}^2$ throat, and the instability is a
violation of the throat's Breitenlohner--Freedman bound. The onset
temperature vanishes at $\astar$ with an essential singularity of
Berezinskii--Kosterlitz--Thouless type. For two-derivative Einstein--Yang--Mills duals in $d = 4$, 5 and 6
dimensions the normal phase has a charged instability if and only if
$C_J/C_T < 8/(d^2(d+1))$, one tenth in $d = 4$. Above $\alpha_L = 0.372$
the gravitational attraction between vortices makes the uniform lattice
unstable and the transition first order, so $B_c$ is only a spinodal;
whether a condensed phase survives above $\astar$ is left open.}


\begin{document}
\maketitle

\section{Introduction}
\label{sec:1}

A strongly coupled plasma with a non-abelian global symmetry can be
magnetically unstable. In the simplest holographic model, Einstein--$SU(2)$
Yang--Mills theory in $\mathrm{AdS}_5$, a magnetic field in the abelian
direction of the gauge group makes the charged vector bosons tachyonic in
their lowest Landau level. This is the holographic form of the
Nielsen--Olesen instability~\cite{Nielsen:1978rm}. In the probe limit, where
the gauge field does not deform the geometry, the plasma condenses once the
field exceeds $B_c \simeq 50\,T^2$~\cite{Ammon:2011je}. The condensate
carries the flux in a lattice of vortices, like the Abrikosov lattice of a
type II superconductor~\cite{Abrikosov:1956sx,Kleiner:1964zz}, and numerical
comparisons of candidate lattices found the triangular one
lowest~\cite{Bu:2012mq}. We show in appendix~\ref{app:probe} that this is a
theorem: the quartic vortex interaction of the probe theory is completely
monotone in the momentum transfer, and Montgomery's
theorem~\cite{Montgomery:1988mt,Cohn:2007universally,Betermin:2016theta}
then makes the triangular lattice the minimum among all one-flux-quantum
Bravais lattices.

All of this ignores the gravity of the field itself. The field sources the
metric with a strength set by the single coupling of the theory,
$\alpha = \kappa^2/\hat g^2$, and enters the geometry through
$\beta = \alpha B^2$. Since $B_c \simeq 5$ in horizon units, $\beta$ is of
order one already at couplings of a few percent: $\beta = 1$ at the onset
when $\alpha = 0.028$. This paper asks what happens once the field bends
the geometry. The answer is that gravity can remove the linear instability
altogether.

The calculation is exact in the coupling and expands only in the
condensate, which is small near the onset. This is possible because the
condensate-free background at any $\beta$ is known: it is the magnetic
brane of D'Hoker and Kraus~\cite{DHoker:2009mmn,DHoker:2009ixq}. The onset
becomes a Sturm--Liouville problem on the brane, and the quartic term a
boundary value problem for the brane's linear response.

The central result is that the critical field rises with the coupling and
diverges at
\begin{equation}
\alpha_\star = \tfrac23~,
\end{equation}
above which the normal phase of the magnetised plasma has no linear
instability in the charged sector, at any field or temperature. The reason
is simple. At zero temperature the interior of the brane is an
$\mathrm{AdS}_3 \times \mathbb{R}^2$ throat in which the flux plane
stretches in proportion to $B$~\cite{DHoker:2009mmn}. The field measured
in throat units, and with it the tachyonic mass of the lowest Landau level,
is therefore a number fixed by $\alpha$, and the onset condition becomes
the Breitenlohner--Freedman
bound~\cite{Breitenlohner:1982bm,Breitenlohner:1982jf} of the
$\mathrm{AdS}_3$ factor. The
tachyon violates it exactly when $\alpha < 2/3$.

Violating the bound is sufficient for the instability; a node count
shows that here it is also necessary, and a positivity argument extends
the result to the whole charged sector. At finite temperature we establish the result
numerically. In $d + 1$ bulk dimensions the bound is
$\alpha_\star(d) = 16/(d(d-1)(d-2))$, and it is the critical coupling for
$d = 4, 5, 6$. Since $\alpha$ is proportional to the ratio of the current
and stress-tensor central charges of the boundary theory, the criterion can
be stated in boundary terms: for $d = 4, 5, 6$ a magnetised plasma with an
Einstein--Yang--Mills dual has a charged instability at zero temperature
if and only if
\begin{equation}
\frac{C_J}{C_T} < \frac{8}{d^2(d+1)}~,
\end{equation}
which in four dimensions says that the current carries less than one tenth
of the stress-tensor central charge. The onset temperature vanishes at
$\alpha_\star$ with an essential singularity of
Berezinskii--Kosterlitz--Thouless type, whose coefficient we derive, though it sets in only at extremely low
temperatures.

\begin{figure}[t]
\centering
\includegraphics{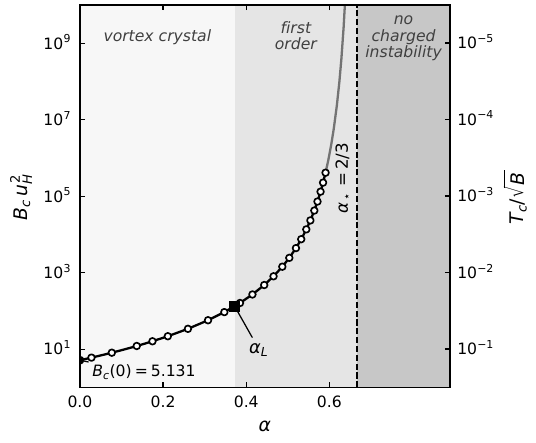}
\caption{The critical field $B_c(\alpha)$ at fixed temperature, on a
logarithmic axis: the collocation ladder (circles) and, beyond its top rung
at $\alpha = 0.591$, direct solves by shooting from the horizon (grey). The
curve rises convexly and has a vertical asymptote at the critical coupling
$\astar = 2/3$. The uniform vortex crystal forms in a second-order
transition for $\alpha < \alpha_L$; for $\alpha_L < \alpha < \astar$ the
transition is first order and $B_c$ is a spinodal; for $\alpha > \astar$
the normal phase has no linear charged instability. The right axis reads
the same curve as the onset temperature at fixed field,
$T_c/\sqrt B = 1/(\pi\sqrt{B_c u_H^2})$, which falls from $0.141$ in the
probe limit to zero at $\astar$.}
\label{fig:1}
\end{figure}

Below $\alpha_\star$ gravity adds an attractive channel to the interaction
between vortices: the exchange of the plasma's stress tensor between
density modulations of the condensate. The attraction barely changes the
shape of the lattice, but it decides its stability. At
\begin{equation}
\alpha_L = 0.372
\end{equation}
the quartic coefficient of the uniform lattice changes sign, the triangular
lattice becomes unstable to long-wavelength density modulation, and the
transition at $B_c$ turns first order, so $\alpha_L$ is a tricritical
point. The plasma thus passes through three
regimes (figure~\ref{fig:1}): a vortex crystal below $\alpha_L$, a
first-order regime up to $\alpha_\star$, in which $B_c$ is only a spinodal,
and no charged instability beyond. Whether a condensed phase survives above
$\alpha_\star$, reached by a first-order jump, is the main question we
leave open.

Einstein--$SU(2)$ Yang--Mills holography goes back
to~\cite{Gubser:2008zu,Gubser:2008wv}, where a chemical potential drives a
$p$-wave condensate; in the backreacted model of~\cite{Gubser:2008zu} the
critical temperature was found numerically to vanish at a critical
coupling. The magnetic brane and its throat are due to D'Hoker and
Kraus~\cite{DHoker:2009mmn,DHoker:2012rlj}, who computed its correlators
with Shah~\cite{DHoker:2010xwl} and also found a
quantum critical point on the charged brane, tuned by field over charge
density at fixed Chern--Simons coupling~\cite{DHoker:2010zpp,DHoker:2010onp}.
The stability of the throat within $D = 5$ $SO(6)$ gauged supergravity,
which uplifts to type IIB on $S^5$, was examined
in~\cite{Almuhairi:2010rb,Almuhairi:2011ws,Donos:2011pn}. For the
equal-charge embedding of D'Hoker and Kraus the scalars of the truncation
are stable~\cite{Almuhairi:2010rb}, but the $SO(6)$ $W$ bosons have a
Nielsen--Olesen mode that violates the $\mathrm{AdS}_3$
bound~\cite{Donos:2011pn},\footnote{The mode was first found in v1
of~\cite{Almuhairi:2011ws}.} and, including their mixing with
the charged scalars, \cite{Donos:2011pn} mapped the unstable embeddings and
found only a narrow window of non-supersymmetric throats with no known
instability next to the supersymmetric ones. In all of these
the coupling is fixed by supergravity and the $U(1)$ embedding is the
parameter. Here the embedding is fixed and the coupling is the parameter,
which is what produces a critical value. In $d = 3$,
\cite{Wong:2013rda} derived the corresponding bound on the magnetic
$\mathrm{AdS}_4$ brane, which our general-$d$ formula reproduces, computed
the onset at finite coupling and temperature, and studied the lattice in
the probe limit. In $d = 3$ we find that the bound is not the critical
coupling (section~\ref{sec:4.3}): a bound state outside the $\mathrm{AdS}_2$
throat keeps the plasma unstable up to $\alpha = 3.037$ rather than $8/3$.

Backreacted vortex lattices of abelian condensates have been constructed
near onset~\cite{Bao:2013fda}, in a circular-cell
approximation~\cite{Tallarita:2019amp} and in full~\cite{Donos:2020viz},
which finds the triangular lattice preferred; the onset on a backreacted magnetic brane
has been studied for an abelian $s$-wave condensate~\cite{Chernicoff:2024ymn},
and a single backreacted vortex with non-abelian orientational moduli was
constructed in~\cite{Tallarita:2019czh}. With an adjoint Higgs field the
$SU(2)$ theory on the magnetic $\mathrm{AdS}_4$ brane instead forms a
monopole wall through a first-order transition~\cite{Bolognesi:2010nb}. The
flat-space vortex lattices of the gluon~\cite{Ambjorn:1980ms} and
electroweak $W$~\cite{Ambjorn:1988fx,Ambjorn:1988tm}, and the proposed
$\rho$-meson condensate of QCD~\cite{Chernodub:2010qx,Adhikari:2024bfa},
are the same mechanism without gravity; with gravity, the same instability
appears near the horizon of a magnetically charged black
hole~\cite{Lee:1991qs,Maldacena:2020skw}. The $\rho$-meson onset has also
been found holographically in the Sakai--Sugimoto
model~\cite{Callebaut:2011ab,Callebaut:2013wba}. Earlier holographic vortex lattices in the probe
limit are~\cite{Maeda:2009vf,Murray:2011gr}. The holographic
Berezinskii--Kosterlitz--Thouless scaling we find is that
of~\cite{Kaplan:2009kr,Jensen:2010ga,Iqbal:2010eh}, with a bulk coupling
tuned through the bound rather than a filling fraction, an
operator dimension, a field~\cite{Faulkner:2010gj} or a flavour
number~\cite{Alho:2013dka}.

Section~\ref{sec:2} sets up the theory and the expansion,
section~\ref{sec:3} gives the onset curve $B_c(\alpha)$,
section~\ref{sec:4} derives the critical coupling, and section~\ref{sec:5}
treats the lattice. Section~\ref{sec:6} lists open questions.
\section{Setup: one coupling, one expansion}
\label{sec:2}

\subsection{The theory has one coupling, and it is a boundary observable}
\label{sec:2.1}

We consider Einstein gravity with a negative cosmological constant coupled to
$SU(2)$ Yang--Mills theory in five dimensions,
\begin{equation}
S = \frac{1}{2\kappa^2}\int d^5x\sqrt{-g}\,\Big[R + 12 - \frac{\alpha}{2}\,F^a_{MN}F^{a\,MN}\Big]~,
\end{equation}
where we have set the AdS radius to one and absorbed the gauge coupling into
the field, $F^a = dA^a + \tfrac12\epsilon^{abc}A^b\wedge A^c$, so that the
Yang--Mills term in its usual normalisation, $-F^2/4\hat g^2$, appears with
the single dimensionless coefficient
\begin{equation}
\alpha = \frac{\kappa^2}{\hat g^2}~.
\end{equation}
This is the only parameter of the theory.\footnote{The backreacted $p$-wave
superfluid of~\cite{Ammon:2009xh} uses the same theory with the unsquared
ratio $\kappa/\hat g$ as its parameter, so its couplings are the square roots
of ours.} In the probe limit $\alpha \to 0$
the gauge field does not deform the geometry, and the calculation reduces to
the one of~\cite{Ammon:2011je}. The equations of motion in Ricci form are
\begin{equation}
R_{MN} + 4\,g_{MN} = \alpha\Big(F^a_{MP}F^{a\phantom{N}P}_{N} - \tfrac16\,g_{MN}F^a_{PQ}F^{a\,PQ}\Big)~, \qquad
D_M F^{a\,MN} = 0~,
\end{equation}
with $D_M$ the gauge-covariant derivative. Since $\alpha$ multiplies the
whole Yang--Mills sector, a background magnetic field $B$ enters the geometry
only through the combination $\beta = \alpha B^2$, and it is this
combination that organises the calculation. Although $\alpha$ is a ratio of
bulk normalisations, it is a boundary observable, proportional to the ratio
$C_J/C_T$ of the current and stress-tensor central charges
(section~\ref{sec:4.4}).

\subsection{The background is exact, and only the condensate is expanded}
\label{sec:2.2}

Near the onset the condensate is small, so a Landau expansion in it is
the right tool as long as the transition is second order; the expansion
itself shows where it is not (section~\ref{sec:5.3}). The backreaction of
the field is not small: $\beta = \alpha B_c^2$ reaches one already at $\alpha \approx 0.03$.
But it need not be treated perturbatively, because the condensate-free
background at any field and temperature can be constructed
non-perturbatively. It is the
magnetic black brane of D'Hoker and Kraus~\cite{DHoker:2009mmn,DHoker:2009ixq},
\begin{equation}
ds^2 = -U\,dt^2 + \frac{dr^2}{U} + e^{2V}\big(dx^2 + dy^2\big) + e^{2Z}dz^2~, \qquad
A^3 = B\,x\,dy~,
\end{equation}
with the flux in the $(x,y)$ plane, which is warped by $V$, and the field
axis along $z$, which is warped separately by $Z$.\footnote{D'Hoker and Kraus write $W$ for our $Z$; we reserve $W$ for the charged gauge field.}
The background field lies in the abelian subgroup generated by $\tau^3$,
so without a condensate the $SU(2)$ theory reduces to Einstein--Maxwell
theory and the brane is exactly the solution of~\cite{DHoker:2009mmn}. The
non-abelian structure enters only through the charged components
$W = A^1 + iA^2$, which see $A^3$ as a $U(1)$ field of unit charge. The
three metric functions obey reduced Einstein--Yang--Mills equations in
which $\alpha$ and $B$ appear only as $\beta$
(appendix~\ref{app:numerics-brane}), so the backgrounds form a
one-parameter family.

We therefore treat the onset, a linear problem on this brane, exactly in
$\beta$, and the lattice perturbatively in the condensate amplitude alone.
Every quantity is then exact in the coupling, at whatever order in the
condensate it is computed. We construct the brane numerically
(appendix~\ref{app:numerics-brane}); appendix~\ref{app:dictionary} gives
the dictionary to the conventions of~\cite{DHoker:2009mmn}.

We work in a fixed-temperature frame throughout: the horizon is at radius
$r_p = 1$ with $U'(1) = 4$, so that $T = 1/\pi$ exactly, and the boundary
normalisation $U, e^{2V}, e^{2Z} \to r^2$ as $r \to \infty$ fixes the
remaining freedom. We quote fields as the dimensionless
$B u_H^2 = B/(\pi T)^2$, where $u_H = 1/\pi T$ is the horizon radius in the
coordinate $u = 1/r$ of the probe literature, so that the probe value
$B_c u_H^2 = 5.13$, that is $B_c \simeq 50\,T^2$, refines the value $5.15$
quoted in~\cite{Ammon:2011je}. At $\beta = 0$ the brane is
AdS$_5$--Schwarzschild, and as $\beta \to \infty$ its interior develops the
long throat of section~\ref{sec:4}.

Table~\ref{tab:notation} collects the notation used in the rest of the
paper.

\begin{table}[t]
\centering
\small
\begin{tabular}{l>{\raggedright\arraybackslash}p{0.62\textwidth}l}
\toprule
symbol & meaning & defined in \\
\midrule
$\alpha$, $\beta$ & coupling $\kappa^2/\hat g^2$; backreaction parameter $\alpha B^2$ & \S\ref{sec:2.1} \\
$U$, $V$, $Z$ & metric functions of the brane; $e^{2V}$ is the proper size of the flux plane & \S\ref{sec:2.2} \\
$B_c u_H^2$ & critical field in horizon units, $B_c/(\pi T)^2$ & \S\ref{sec:2.2} \\
$W$, $w$, $w_0$ & charged field $A^1 + iA^2$; its radial profile; the zero mode at onset & \S\ref{sec:3.1} \\
$P$, $Q$ & Sturm--Liouville coefficients $Ue^Z$, $Be^{Z-2V}$ (written $p_2$, $Bp_1$ in \S\ref{sec:5}) & \S\ref{sec:3.1} \\
$\alpha_\star$, $\alpha_c(3)$ & critical coupling $2/3$; its $d = 3$ counterpart $3.037$ & \S\S\ref{sec:4.2}, \ref{sec:4.3} \\
$\ell_3$, $v$, $b$, $b_h$ & throat radius; frozen $e^{2V}$; throat field $B\ell_3^2e^{-2V}$; $b$ at the horizon & \S\S\ref{sec:4.1}, \ref{sec:4.2} \\
$\varepsilon$ & distance of the horizon invariant from the fixed point, $6 - \beta e^{-4V}$ & \S\ref{sec:4.2} \\
$\rho$, $h$, $\rho_\star$ & radius $r - r_0$ in the throat; BTZ horizon; doubling radius of $e^{2V}$ & \S\S\ref{sec:4.1}, \ref{sec:4.2}, \ref{sec:4.6} \\
$c_0$, $c_1$ & coefficients of the threshold mode, $\rho^{(d-2)/2}w = c_0 + c_1\ln(\rho/\rho_\star)$ & \S\ref{sec:4.2} \\
$\nu$, $\chi$, $\phi_\star$ & $\sqrt{b_h - 1}$; horizon and outer phases & \S\ref{sec:4.6} \\
$C_J$, $C_T$ & current and stress-tensor central charges & \S\ref{sec:4.4} \\
$\eta$, $\epsilon$ & condensate amplitude; coefficient of $\eta^2$, $\propto B - B_c$ & \S\ref{sec:5.1} \\
$G$, $s$, $\gamma$ & reciprocal lattice vector; squared harmonic $G^2$; $s/B$ & \S\ref{sec:5.1} \\
$K$, $K_0$, $K_{G=0}$ & quartic kernel; its limit $s \to 0$; the uniform response & \S\ref{sec:5.1} \\
$C_4$, $X$, $\mathcal{P}$ & contact term; photon screening; metric pairing & \S\ref{sec:5.1} \\
$\tau$, $C(\tau)$ & lattice shape; shell sum selecting it ($C_\triangle$, $C_\square$ at $e^{i\pi/3}$, $i$) & \S\S\ref{sec:5.1}, \ref{sec:5.2} \\
$\alpha_L$, $\alpha_I$ & uniform lattice unstable; $C$ unbounded towards the stripe edge & \S\S\ref{sec:5.3}, \ref{sec:5.2} \\
\bottomrule
\end{tabular}
\caption{Notation. Symbols used only in the appendices are defined there.}
\label{tab:notation}
\end{table}
\section{The critical field rises convexly with the coupling}
\label{sec:3}

\subsection{The onset is a Sturm--Liouville problem on the brane}
\label{sec:3.1}

In the probe limit the instability is the one of Nielsen and Olesen: the
charged combination $W = A^1 + iA^2$ of the gauge field is a vector boson of
gyromagnetic ratio $g = 2$, and in the lowest Landau level its
magnetic-moment coupling to the background field is strong enough to make it
tachyonic. As shown in~\cite{Ammon:2011je}, the mode that condenses first is
the polarised lowest Landau level
\begin{equation}
W_x = w(r)\,\psi_0(x,y)~, \qquad W_y = -i\,W_x~,
\end{equation}
where $\psi_0$ is any function in the lowest Landau level of the charge-one
field $A^3 = Bx\,dy$, and $w(r)$ is the radial profile. In the boundary
theory what condenses is the charged spatial current
$\langle J^1_i \pm iJ^2_i\rangle$, modulated in the plane. The same
instability for a charged vector of general mass and gyromagnetic ratio,
which reduces to the $SU(2)$ case at zero mass and $g = 2$, was studied in
the probe limit in~\cite{Cai:2013pda,Cai:2013kaa}. In the gauge $W_r = 0$,
$W_t$ and $W_z$ vanish identically for this mode, and on the magnetic brane
the same ansatz reduces the linearised Yang--Mills equations exactly to a
Sturm--Liouville problem in $r$ with the field as the eigenvalue,
\begin{equation}
\big(P\,w'\big)' + Q\,w = 0~, \qquad P = U\,e^{Z}~, \qquad Q = B\,e^{Z-2V}~.
\end{equation}
One feature of this equation matters for what follows. On the polarised lowest Landau level
the transverse kinetic term of $W$ vanishes identically, since
$F^W_{xy} = 0$ for this polarisation, so the only term without a radial
derivative is the non-abelian magnetic-moment coupling. This is the $g = 2$
mechanism, and it carries a single factor $e^{-2V}$: the moment term samples
the warping of the flux plane and nothing else, which is what makes the
field drop out in the throat (section~\ref{sec:4.2}). The onset is
determined by this one equation on the full brane (the truncation is
consistent: at linear order in $w$ nothing sources the metric, the abelian
field $A^3$ or the other colour components;
appendix~\ref{app:charged-zero-mode}). In the probe limit $\beta \to 0$ the equation
reduces to the probe problem, $P \to f/u$ and $Q \to B/u$ in the coordinate
$u = 1/r$, with $f = 1 - u^4$ the blackening factor of
$\mathrm{AdS}_5$--Schwarzschild, and the same critical field $B_c u_H^2 = 5.13126764$.

The boundary conditions are those of the probe calculation, regularity at
the horizon and no source at the boundary, and the brane's logarithmic
asymptotics do not obstruct them (appendix~\ref{app:charged-zero-mode}).
Since $\psi_0$ is
arbitrary within the lowest Landau level, the onset is infinitely
degenerate, and the lattice is selected only at quartic order, which is the
subject of section~\ref{sec:5}.

\subsection{The onset curve is a single sweep in \texorpdfstring{$\beta$}{beta}}
\label{sec:3.2}

At fixed $\alpha$ the critical field is the smallest eigenvalue $B$ of the
Sturm--Liouville problem on the brane at the same $B$. This looks like a
self-consistency problem, but it needs no iteration: the background depends
on $\alpha$ and $B$ only through $\beta$, and at fixed $\beta$, $B$ enters
the operator only as the eigenvalue. We therefore construct the brane at a given
$\beta$, solve for its lowest
eigenvalue $B_{\rm eig}(\beta)$, and read off the coupling at which this
brane is the self-consistent onset,
\begin{equation}
B_c = B_{\rm eig}(\beta)~, \qquad \alpha(\beta) = \frac{\beta}{B_{\rm eig}(\beta)^2}~.
\end{equation}

\subsection{Backreaction stabilises the normal phase without saturating}
\label{sec:3.3}

The critical field at fixed temperature is shown in figure~\ref{fig:1}, in
units of the horizon radius, that is $B_c/T^2 = \pi^2 B_c u_H^2$. We solved
the sweep on a dense grid in $\beta$ up to $\beta = 45$, where
$\alpha = 0.1745$ and $B_c u_H^2 = 16.06$, and on a sparse sequence of 23
values from $\beta = 1$ to $10^{11}$, which we call the ladder and its
members rungs; the rungs beyond $\beta = 45$ are used for the approach
to $\alpha_\star$ in section~\ref{sec:4.6}. The curve is
monotone and convex, and its tangent at $\alpha = 0$ is already ten percent
off at $\alpha \approx 0.08$. Backreaction thus stabilises the normal phase: at a
given temperature the magnetised plasma needs a larger field to condense
when the field is allowed to deform the geometry, and the effect grows
faster than linearly. That backreaction works against the instability, and
removes it at zero temperature beyond a critical coupling, was found
in~\cite{Wong:2013rda} in four bulk dimensions; the exact curve shows how
$B_c$ diverges on the approach to $\alpha_\star$. Along the ladder $B_c$ rises by more than three orders
of magnitude while $\alpha$ moves from $0.4$ to $0.59$,
which is the first sign that the curve has a vertical asymptote.
\section{No charged instability above \texorpdfstring{$\alpha_\star = 2/3$}{alpha* = 2/3}}
\label{sec:4}

\subsection{At zero temperature the brane interior is \texorpdfstring{$\mathrm{AdS}_3 \times \mathbb{R}^2$}{AdS3 x R2}}
\label{sec:4.1}

Since the field is measured in units of $T^2$, cooling at fixed physical
field sends $B u_H^2$, and with it $\beta = \alpha (B u_H^2)^2$, to
infinity: the zero-temperature limit at fixed $\alpha$ is
$\beta \to \infty$. In this limit the interior of the brane develops a long
throat~\cite{DHoker:2009mmn}. It approaches the product geometry
$\mathrm{AdS}_3 \times \mathbb{R}^2$, in which the flux plane $(x,y)$ has
frozen at a finite proper size and the direction $z$ along the field has
joined $t$ and $r$ to form a three-dimensional anti-de Sitter space. In our
conventions $U = (r - r_0)^2/\ell_3^2$, $e^{2Z} = c_Z (r - r_0)^2$,
$e^{2V} = v$ solves the full Einstein--Yang--Mills equations if and only
if
\begin{equation}
\label{eq:throat}
\ell_3^2 = \tfrac13, \qquad v = B\sqrt{\alpha/6}~.
\end{equation}
The constants $c_Z$ and $r_0$ can be removed by rescaling $z$ and shifting
$r$.

\begin{figure}[t]
\centering
\includegraphics{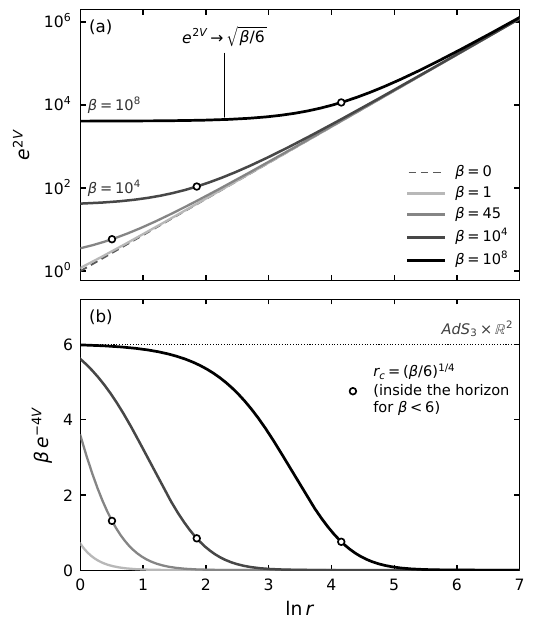}
\caption{The magnetic brane forms its throat. (a)~The proper size $e^{2V}$
of the flux plane against $\ln r$ at fixed temperature, for
$\beta = 0, 1, 45, 10^4, 10^8$ (light to dark; $\beta = 0$ dashed). Outside
the crossover radius $r_c$ (circles) it grows as $r^2$, as in
$\mathrm{AdS}_5$; inside it the flux plane freezes at
$e^{2V} \to v = \sqrt{\beta/6}$. (b)~The invariant $\beta e^{-4V}$, which
is $0$ at the boundary and $6$ on the $\mathrm{AdS}_3 \times \mathbb{R}^2$
fixed point. Its horizon value is $3.58$, $5.613$ and $5.9877$ at
$\beta = 45$, $10^4$ and $10^8$, and the plateau near $6$ lengthens as
$\ln r_c = \tfrac14\ln(\beta/6)$.}
\label{fig:throat}
\end{figure}

Neither $v$ nor $B$ is meaningful on its own, since rescaling the flux
plane, $(x,y) \to k\,(x,y)$, sends both $e^{2V}$ and $B$ to themselves
divided by $k^2$. The scale-invariant form of~\eqref{eq:throat} is
\begin{equation}
\alpha B^2 e^{-4V} = 6, \qquad B\,e^{-2V} = \sqrt{6/\alpha}~.
\end{equation}
The first is the condition for the throat to exist. The second is the
magnetic field measured in the proper units of the frozen flux plane, and
it controls the condensate. At finite $\beta$ the throat joins the
$\mathrm{AdS}_5$ region at the crossover radius $r_c = (\beta/6)^{1/4}$,
where the outer growth $e^{2V} \simeq r^2$ meets the frozen value
$\sqrt{\beta/6}$ (figure~\ref{fig:throat}).

At small temperature the throat is a BTZ black hole~\cite{Banados:1992wn}
times the flux plane:
$U = ((r-r_0)^2 - h^2)/\ell_3^2$, with horizon at $r - r_0 = h$, solves the
same equations with the same $\ell_3$ and $v$
(appendix~\ref{app:numerics-brane}). This is the throat on the
finite-temperature onset curve, and it governs the approach to
$\alpha_\star$ in section~\ref{sec:4.6}.

\subsection{The field drops out and the onset becomes a Breitenlohner--Freedman bound}
\label{sec:4.2}

On the throat metric the Sturm--Liouville equation of section~\ref{sec:3}
becomes an Euler equation in $\rho = r - r_0$, $(\rho^3 w')' + b\,\rho\,w = 0$, and $w = \rho^{-\Delta}$
gives the indicial equation
\begin{equation}
\Delta(\Delta - 2) = -b, \qquad b \equiv B\,\ell_3^2\,e^{-2V} = \sqrt{\frac{2}{3\alpha}}~.
\end{equation}
The field has cancelled because the frozen flux plane grows in
proportion to $B$, so the field in throat units, $b$, is fixed by the
coupling alone. In the language of the $\mathrm{AdS}_3$ factor, the lowest
Landau level of the $W$ boson is a scalar of mass $m^2\ell_3^2 = -b$, and
the Breitenlohner--Freedman
bound~\cite{Breitenlohner:1982bm,Breitenlohner:1982jf}
$m^2\ell_3^2 \ge -1$ is violated exactly
when $b > 1$. Violating it is sufficient for an instability at zero
temperature, and we show below that it is also necessary, so that
\begin{equation}
\text{the instability survives to } T = 0 \quad\Longleftrightarrow\quad \alpha < \alpha_\star = \tfrac23~.
\end{equation}
We call $\alpha_\star$ the critical coupling. As $\alpha \to 0$, $b \to \infty$
and we recover the probe tachyon of~\cite{Ammon:2011je}. Increasing
$\alpha$ lowers the throat field $b$, which squeezes the tachyon until, at
$\alpha_\star$, the two indicial roots meet at $\Delta = 1$.

That the bound is also necessary needs an argument, because
violating the bound of a throat is sufficient for an instability but not in
general necessary~\cite{Faulkner:2010gj,Horowitz:2010gk}: a bound state can
sit in the crossover between the throat and the boundary. For charged
scalars the two criteria were found numerically to
agree~\cite{Denef:2009tp}, and the bound is used as the criterion
in~\cite{Gubser:2009cg,Horowitz:2009ij}, but this is a
property of each model, and in $d = 3$ they disagree
(section~\ref{sec:4.3}).

Here the zero-temperature brane is a single geometry, and the coupling
enters the zero-mode problem only through $b = \sqrt{\alpha_\star/\alpha}$,
which plays the role of an eigenvalue. In the throat the mode behaves as
$(r - r_0)^{-1 \pm \sqrt{1 - b}}$, so the operator has a continuum for
$b > 1$, and a zero mode at a coupling above $\alpha_\star$ would be a
bound state below the threshold $b = 1$. Such a bound state and the
source-free solution at threshold have vanishing Wronskian at both ends,
at the boundary because both are source-free and in the throat because the
bound state decays, so by Sturm's comparison
theorem~\cite[theorem~9.39]{Teschl:2009mmqm} the threshold solution would
have a zero. At threshold the solution in the throat has the form
$(r - r_0)\,w = c_0 + c_1\ln\big((r - r_0)/\rho_\star\big)$, with
$\rho_\star$ the top of the throat (section~\ref{sec:4.6}), and a zero
inside the throat would show up as $c_1/c_0 > 0$. We find
no zero and $c_1/c_0 = -0.035$: the would-be zero lies far
outside the throat, at $\ln\big((r - r_0)/\rho_\star\big) = +28.6$. There is therefore
no bound state, and for every $\alpha > \alpha_\star$ the D'Hoker--Kraus
brane is stable against this mode at every field. A direct check agrees:
integrated from the throat to the boundary at couplings between
$\alpha_\star$ and $4\alpha_\star$, the solution that decays in the throat
always arrives with a source of the same sign
(appendix~\ref{app:numerics-onset}).

At finite temperature the statement is numerical. On the onset curve
$\alpha/\alpha_\star = (1 - \varepsilon/6)/b_h^2$, with
$\varepsilon = 6 - \beta e^{-4V}$ and $b_h = B_c\ell_3^2 e^{-2V}$ at the
horizon. Since $\varepsilon > 0$ at finite temperature, $b_h \ge 1$
implies $\alpha < \alpha_\star$. Shooting from the horizon up to
$\beta \approx 10^{66}$, where $\alpha = 0.6626$, the onset coupling rises
strictly and stays below $\alpha_\star$, with $b_h \ge 1.003$
(appendix~\ref{app:numerics-onset}); beyond that the matching analysis of
section~\ref{sec:4.6} takes over.

\subsection{In \texorpdfstring{$d+1$}{d+1} dimensions \texorpdfstring{$\alpha_\star = 16/(d(d-1)(d-2))$}{alpha* = 16/(d(d-1)(d-2))}}
\label{sec:4.3}

The argument is not specific to four boundary dimensions. The reduction to
the polarised lowest Landau level goes through as before, with
$P = U e^{(d-3)Z}$ and $Q = B e^{(d-3)Z - 2V}$; the $d - 3$ directions along
the field only contribute a volume factor. The throat is now
$\mathrm{AdS}_{d-1} \times \mathbb{R}^2$, one of the magnetic vacua that
exist in every dimension~\cite{Lu:2013eoa}, with
\begin{equation}
\ell_{d-1}^2 = \frac{(d-2)^2}{d(d-1)}, \qquad \frac{B\ell_{d-1}^2}{v} = \frac{(d-2)^{3/2}}{\sqrt{\alpha\,d(d-1)}}~,
\end{equation}
and the indicial equation becomes $\Delta(\Delta - (d-2)) = -B\ell_{d-1}^2/v$,
with $v = e^{2V}$ again the frozen size of the flux plane. Comparing with
the Breitenlohner--Freedman value $(d-2)^2/4$ for $\mathrm{AdS}_{d-1}$
gives
\begin{equation}
\alpha_\star(d) = \frac{16}{d(d-1)(d-2)}~,
\end{equation}
which is $8/3$, $2/3$, $4/15$ and $2/15$ for $d = 3, 4, 5, 6$.

For $d = 4, 5, 6$ this bound is the critical coupling. We constructed the
zero-temperature brane in each dimension, flowing from
$\mathrm{AdS}_{d+1}$ into the throat along its single irrelevant
deformation, and the threshold mode on it has no zero, with
$c_1/c_0 < 0$ (table~\ref{tab:dims}). At finite temperature the
onset coupling for $d = 5, 6$, as for $d = 4$, rises monotonically, stays
below $\alpha_\star(d)$ and approaches it as $T \to 0$
(appendix~\ref{app:numerics-onset}).

In $d = 3$ the bound is not the critical coupling. The throat is
$\mathrm{AdS}_2 \times \mathbb{R}^2$, and on the extremal magnetic
Reissner--Nordstr\"om--$\mathrm{AdS}_4$ brane the threshold mode has one
zero deep in the throat, so a bound state sits in the crossover below the
$\mathrm{AdS}_2$ bound. It disappears only at
\begin{equation}
\alpha_c(3) = 3.0370~,
\end{equation}
which we find by three methods and confirm with an independent
construction (appendix~\ref{app:numerics-onset}). At finite temperature the
onset coupling rises through $8/3$ and tends to $\alpha_c(3)$ as $T \to 0$.
The bound of~\cite{Wong:2013rda} on the magnetic $\mathrm{AdS}_4$ brane,
$\alpha_\star(3) = 8/3$ in our variables, is therefore sufficient for the
instability but, contrary to the statement in~\cite{Wong:2013rda} that the
$W$ bosons condense only for $\alpha < 8/3$, not necessary; there the
zero-temperature end of the onset curve was taken from the bound. Our
finite-temperature onset crosses $8/3$ only at $T/\sqrt B = 7.4\times10^{-4}$.

\subsection{The criterion is \texorpdfstring{$C_J/C_T < 8/(d^2(d+1))$}{C\_J/C\_T < 8/(d\^{}2(d+1))}}
\label{sec:4.4}

The coupling $\alpha = \kappa^2/\hat g^2$ depends on how the Yang--Mills term
is normalised relative to the Einstein--Hilbert term, and papers differ;
\cite{Wong:2013rda}, for instance, uses $2e^2L^2/\kappa^2$. A
convention-free form uses the central charges $C_T$ and $C_J$, which
normalise the two-point functions of the stress tensor and of the current
that the magnetic field couples to; criteria of this kind were written in
terms of $C_T$ and $C_J$ in~\cite{Nakayama:2015hga}, as an AdS form of the
weak gravity conjecture, for the $\mathrm{AdS}_2$ bound of the electric
brane. In the conventions
of~\cite{Osborn:1993cr}, Einstein gravity with a Maxwell term
$-F^2/4\hat g^2$ in $\mathrm{AdS}_{d+1}$ gives~\cite{Buchel:2009sk,Freedman:1998tz}
\begin{equation}
C_T = \frac{d+1}{d-1}\,\frac{\Gamma(d+1)}{\pi^{d/2}\,\Gamma(d/2)}\,\frac{L^{d-1}}{\kappa^2}, \qquad
C_J = \frac{(d-2)\,\Gamma(d)}{2\pi^{d/2}\,\Gamma(d/2)}\,\frac{L^{d-3}}{\hat g^2}~,
\end{equation}
where for the non-abelian field the second formula holds for each
generator, in particular for the $\tau^3$ that carries the background flux.
The ratio is
\begin{equation}
\frac{C_J}{C_T} = \frac{(d-1)(d-2)}{2d(d+1)}\,\alpha~,
\end{equation}
which is $3\alpha/20$ in $d = 4$, and the result of section~\ref{sec:4.3}
becomes
\begin{equation}
\text{the magnetised plasma is unstable at } T = 0 \quad\Longleftrightarrow\quad \frac{C_J}{C_T} < \frac{8}{d^2(d+1)}~.
\end{equation}
Here unstable means that the normal phase has a linear charged
instability at some field. $C_J$ is normalised by the algebra of the charges, $[Q^a, Q^b] = i\epsilon^{abc}Q^c$, so
that a doublet carries $T^3 = \pm\frac12$ (appendix~\ref{app:dictionary}).
The criterion holds for $d = 4, 5, 6$; in $d = 3$ the bound value $2/9$ is replaced by
$\alpha_c(3)/12 = 0.253$ (table~\ref{tab:dims}).

\begin{table}[t]
\centering
\small
\setlength{\tabcolsep}{4.5pt}
\begin{tabular}{cccccccc}
\toprule
$d$ & throat & $\alpha_\star(d)$ & $c_1/c_0$ & critical coupling & $C_J/C_T$ threshold & \multicolumn{2}{c}{supersymmetric check} \\
 & & & & & & $\alpha$ & $C_J/C_T$ \\
\midrule
$3$ & $\mathrm{AdS}_2 \times \mathbb{R}^2$ & $8/3$ & $+0.123$ & $\alpha_c(3) = 3.037$ & $0.253$ & $2$ & $1/6$ \\
$4$ & $\mathrm{AdS}_3 \times \mathbb{R}^2$ & $2/3$ & $-0.035$ & $\alpha_\star$ & $1/10$ & $1/4$ & $3/80$ \\
$5$ & $\mathrm{AdS}_4 \times \mathbb{R}^2$ & $4/15$ & $-0.350$ & $\alpha_\star$ & $4/75$ & -- & -- \\
$6$ & $\mathrm{AdS}_5 \times \mathbb{R}^2$ & $2/15$ & $-0.996$ & $\alpha_\star$ & $2/63$ & -- & -- \\
\bottomrule
\end{tabular}
\caption{The critical coupling by boundary dimension $d$: the throat, its
Breitenlohner--Freedman value $\alpha_\star(d) = 16/(d(d-1)(d-2))$, the
logarithmic ratio $c_1/c_0$ of the zero-temperature threshold mode,
$\rho^{(d-2)/2}w = c_0 + c_1\ln(\rho/\rho_\star)$ in the throat (positive signals a bound state in the
crossover), the resulting critical coupling, the threshold in $C_J/C_T$, and
the two supersymmetric theories against which the map is checked
(appendix~\ref{app:dictionary}).}
\label{tab:dims}
\end{table}

The criterion puts a number on the description
in~\cite{Ammon:2009xh,Wong:2013rda} of the coupling as the fraction of the
degrees of freedom that are charged. The one top-down case with a known
answer agrees with it. The $\mathcal{N} = 1$ R-current of any
four-dimensional superconformal theory has $C_R/C_T = 1/10$
(appendix~\ref{app:dictionary}), exactly the threshold for a vector of
unit R-charge. In $\mathcal{N} = 4$ super-Yang--Mills theory in an
equal-charge R-symmetry field, whose throat is that of D'Hoker and Kraus
uplifted on $S^5$, the $SO(6)$ $W$ bosons carry R-charge $4/3$ and so see
$\alpha = \tfrac23\big(\tfrac34\big)^2 = \tfrac38 < \alpha_\star$; their
throat mass $\ell_3^2m^2 = -4/3$, from eq.~(3.21) of~\cite{Donos:2011pn} at equal charges, is $-b$ at this
coupling, and they are the unstable mode found
in~\cite{Almuhairi:2011ws,Donos:2011pn}. That $3/8$ lies close to $\alpha_L$
does not place this theory in either lattice regime of section~\ref{sec:5},
which treats Einstein--$SU(2)$ theory only. In any theory whose magnetic
sector reduces to such a charged vector on an
$\mathrm{AdS}_{d-1} \times \mathbb{R}^2$ throat, $C_J/C_T$ below the
threshold implies an instability. The converse also needs the absence of a
bound state in the crossover, which we have shown for the
Einstein--Yang--Mills brane and which fails in $d = 3$.

We check the map from $\alpha$ to $C_J/C_T$ at two points where both sides
are fixed independently (appendix~\ref{app:dictionary}). In Romans' gauged
supergravity in $d = 4$, $\alpha_{\rm SUSY} = 1/4$, and the
$\mathcal{N} = 2$ Ward identities give $C_J(SU(2)_R)/C_T = 3/80 =
(3/20)(1/4)$. In minimal gauged supergravity in $d = 3$,
$\alpha_{\rm SUSY} = 2$ and $C_R/C_T = 1/6 =
\alpha/12$~\cite{Closset:2012ru}. Both points lie inside the
unstable window, the $d = 4$ one in the crystal regime of
section~\ref{sec:5}. Einstein--$SU(2)$ theory with $SU(2)$ flux is not a
consistent truncation of Romans' theory, since Romans' scalar is sourced
by the flux, so the comparison fixes the map, not the phase diagram of a string
vacuum. Whether any top-down theory reaches $\alpha_\star$ is open.

\subsection{No other charged mode is unstable above \texorpdfstring{$\alpha_\star$}{alpha*}}
\label{sec:4.5}

The critical coupling bounds the whole charged sector, not only the
polarised lowest Landau level for which it was derived. Static,
$z$-independent fluctuations of $W_\mu$ on any diagonal magnetic brane sort
by a single Landau index $n \ge 0$, since the Landau ladder operators shift
the level by one and the metric depends on $r$ only; $n = 0$ is the
polarised lowest Landau level (appendix~\ref{app:charged-channels}). Every
level with $n \ge 1$ is stable: its static energy, a sum of positive
radial terms and a positive semi-definite transverse form, vanishes only
on pure gauge, so no source-free, horizon-regular zero mode exists.
Only $n = 0$ has a negative transverse term, and its competition with the
radial gradient is the eigenvalue problem of section~\ref{sec:3}. The
argument uses only positivity, so it holds at every $\alpha$, $B$ and $T$,
and it survives momentum along $z$.

The argument also covers time-dependent perturbations. The brane has no
$A_t$ and no $g_{ti}$, so a growing mode must have real $\omega^2$, hence
$\omega = i\Gamma$ and negative static energy. An instability can therefore
only set in through a static zero mode at $\omega = 0$
(appendix~\ref{app:charged-channels}). Together with the $n = 0$
result, the charged sector has no linear instability above $\alpha_\star$.

We assume that the neutral sector, $A^3$ and the metric, is stable. No
growing neutral mode of the D'Hoker--Kraus brane is known. None has been
found at zero momentum in the scalar and tensor
channels~\cite{Janiszewski:2015ura,Shah:2024lym}, or, with a Chern--Simons
term, along the field~\cite{Ammon:2017ded}. Static modes across the field
respect the bound in the throat of the $U(1)^3$
truncation~\cite{Donos:2011pn}. In our theory, those of the sector the
condensate sources respect it (appendix~\ref{app:response-throat}) and have
no zero mode on any brane we solved (section~\ref{sec:5}). The other static
sectors, and all time-dependent modes across the field, have not been
computed. The known modulated instabilities of
magnetic branes need couplings our theory does not have: a neutral scalar
coupled to $F_{\mu\nu}F'^{\mu\nu}$, with $F'$ a second abelian field
strength~\cite{Donos:2011qt,Donos:2016hsd}, or, for the Chern--Simons
instability of~\cite{Nakamura:2009tf}, an electric background and a cubic
Casimir, which $SU(2)$ lacks.

\subsection{The approach to \texorpdfstring{$\alpha_\star$}{alpha*} has Berezinskii--Kosterlitz--Thouless form}
\label{sec:4.6}

We continued the onset calculation of section~\ref{sec:3} up the ladder of
section~\ref{sec:3.3} to $\beta = 10^{11}$. The interior reaches
the fixed point, with the invariant $\beta e^{-4V}$ at the horizon rising to
$6$, and the throat field $b_h$ at the horizon falls towards the bound
$1$ (appendix~\ref{app:numerics-onset}). Since
$2/(3b_h^2) = 6\alpha/(\beta e^{-4V})$ on the onset curve, together these
say $\alpha \to 2/3$. The approach is very slow: at $\beta = 10^{11}$,
where $B_c u_H^2 = 4.1 \times 10^5$, the coupling has only reached $0.591$.

This is the holographic Berezinskii--Kosterlitz--Thouless scaling
of~\cite{Kaplan:2009kr,Jensen:2010ga,Iqbal:2010eh,Jensen:2010vx,Evans:2010hi},
a Breitenlohner--Freedman bound crossed as a parameter is tuned. Here it
is the bound of an $\mathrm{AdS}_3$ throat, whereas
\cite{Iqbal:2010eh,Jensen:2010vx,Evans:2010hi} and the D3/D5 system
of~\cite{Jensen:2010ga} cross an $\mathrm{AdS}_2$ bound, and the holographic picture
of~\cite{Kaplan:2009kr} is a scalar below the bound of the full
$\mathrm{AdS}_{d+1}$. With $\nu = \sqrt{b_h - 1}$ the mode in the throat oscillates, up to
corrections of order $\varepsilon$, as
$\rho\,w \propto \cos(\nu\ln\rho + \ldots)$, with $\rho = r - r_0$, and the
onset is where the throat is long enough to hold the solution that joins
the horizon to the ultraviolet without an extra node.

\begin{figure}[t]
\centering
\includegraphics{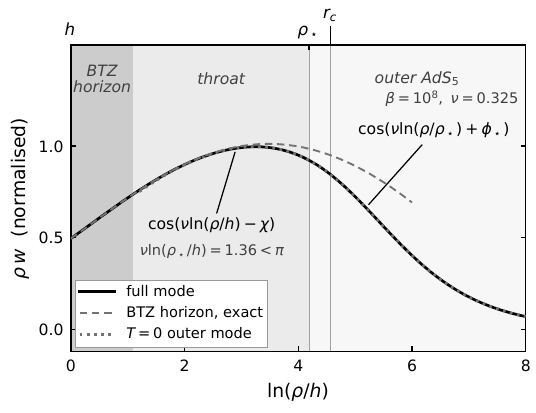}
\caption{The matching diagram: the marginal mode on the $\beta = 10^8$ rung
against $\ln(\rho/h)$, with the exact BTZ horizon solution and the $T = 0$
outer mode laid over it, the three regions shaded and $h$, $\rho_\star$ and
the crossover radius $r_c$ marked.}
\label{fig:matching}
\end{figure}

Both ends of the oscillation can be computed, and matching them has no free
parameter. At the horizon the solution on the BTZ throat is a
hypergeometric function, whose connection formula gives a phase $\chi(\nu)$
in closed form. At the ultraviolet end, the outer region of every rung is,
in scaled variables, the same zero-temperature brane; integrating the
source-free mode inwards from the boundary on it gives a phase
$\phi_\star(\nu)$ in the throat. The two solutions match when
\begin{equation}
\nu\,\ln(\rho_\star/h) = \chi(\nu) + \phi_\star(\nu)~,
\end{equation}
where $h$ is the BTZ horizon and $\rho_\star$ the radius at which the
proper area $e^{2V}$ of the flux plane has doubled from its throat value
(figure~\ref{fig:matching}). This condition reproduces $\nu$ on the ladder
and the direct solves by horizon shooting (appendix~\ref{app:numerics-matching}).

As $\nu \to 0$, $\chi + \phi_\star \to \pi$: the throat holds one
half-oscillation of the mode. With $\ln(\rho_\star/h) \simeq \tfrac14\ln\beta$
(appendix~\ref{app:numerics-matching}),
$\nu^2 \simeq (\alpha_\star - \alpha)/2\alpha_\star$ and
$\beta \propto B^2/T^4$ at fixed field, the onset obeys
\begin{equation}
\ln\beta_c \simeq \frac{8\pi}{\sqrt{3(\alpha_\star - \alpha)}}~, \qquad
T_c \propto \sqrt{B}\,\exp\!\Big(-\frac{2\pi}{\sqrt{3(\alpha_\star - \alpha)}}\Big) = \sqrt B\,e^{-3.63/\sqrt{\alpha_\star-\alpha}}~.
\end{equation}
This is the law of~\cite{Iqbal:2010eh,Jensen:2010vx}; the same law arises
from the annihilation of fixed points, with no bound crossed,
in~\cite{Faedo:2019nxw}. Since $\alpha > \alpha_L$ wherever it applies,
$T_c$ here is the spinodal of the normal phase (section~\ref{sec:5.3}): the
essential singularity is that of the onset of the instability, not of an
infinite-order transition.

\begin{figure}[t]
\centering
\includegraphics{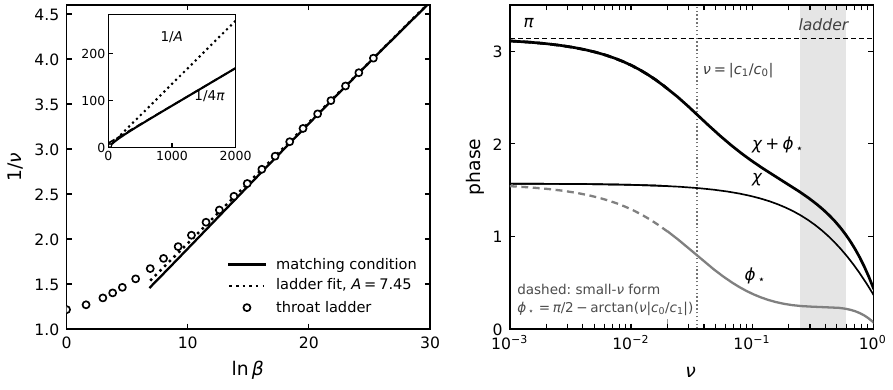}
\caption{The approach law. Left: $1/\nu$ against $\ln\beta$ with the
ladder, the matching prediction, the ladder fit and (inset, to
$\ln\beta = 2000$) the $1/4\pi$ asymptote. Right: $\chi$, $\phi_\star$ and
$\chi + \phi_\star$ against $\nu$ with the $\pi$ limit, the turnover at
$\nu = |c_1/c_0| = 0.035$ and the throat part of the ladder, $0.25 \le \nu \le 0.6$.
Below $\nu = 0.02$ (dashed) $\phi_\star$ is the small-$\nu$ form
$\pi/2 - \arctan(\nu|c_0/c_1|)$ rather than a direct solve, and so is the
approach to $1/4\pi$ in the inset.}
\label{fig:approach}
\end{figure}

The law is asymptotic, and it sets in only at extremely low temperature.
The horizon phase is analytic in $\nu$,
$\chi = \pi/2 - 2\nu\ln 2 + O(\nu^3)$, so its corrections only shift
$\ln\beta_c$ by a constant. The outer phase is not. At $\alpha_\star$ the
mode arrives in the throat almost entirely as the non-logarithmic solution,
with $c_1/c_0 = -0.035$, and matching it onto
$\cos(\nu\ln(\rho/\rho_\star) + \phi_\star)$ gives
$\tan\phi_\star = |c_1/c_0|/\nu + 0.50\,\nu + O(\nu^3)$. So $\phi_\star$
reaches its limit $\pi/2$ only for $\nu \ll 0.035$, that is
$\ln\beta \gg 400$. Over the throat part of the ladder, $\beta \ge 10^3$ or
$0.25 \le \nu \le 0.6$, $\phi_\star$ instead sits on a plateau of about
$0.22$--$0.25$ (figure~\ref{fig:approach}, right). In this range the
throat needs to hold less phase, so the plasma condenses more easily than the asymptotic law
predicts. A direct fit on the ladder gives
\begin{equation}
T_c \propto \sqrt B\,e^{-2.25/\sqrt{\alpha_\star - \alpha}}~,
\end{equation}
a coefficient $38\%$ below the asymptotic one. Shooting from the horizon
far beyond the ladder confirms that the effective coefficient rises
slowly towards its asymptotic value, as the matching condition predicts
(appendix~\ref{app:numerics-matching}); the asymptotic law takes over only
at $T_c/\sqrt B \sim e^{-100}$.

\section{The triangular lattice forms below \texorpdfstring{$\alpha_L = 0.372$}{alpha\_L = 0.372}}
\label{sec:5}

\subsection{Gravity adds an attractive channel to the vortex interaction}
\label{sec:5.1}

Below the critical coupling the normal phase is unstable, and we ask which
lattice forms and whether it is stable. We follow the probe-limit route
of~\cite{Bu:2012mq} and expand the free energy to quartic order in the
condensate, now on the brane and with the condensate sourcing the metric as
well as the abelian field. Just above the onset
$W_x = \eta\,w_0(r)\,\psi(x,y)$, $W_y = -iW_x$, with $w_0$ the zero mode of
section~\ref{sec:3} and $\psi$ a lowest-Landau-level function with one flux
quantum per unit cell. The Fourier coefficients $\omega_G$ of $|\psi|^2$,
with $G$ a reciprocal lattice vector, have $|\omega_G|^2 = e^{-\gamma_G/2}$ with
$\gamma_G = G^2/B$ (appendix~\ref{app:probe-lll-identity}), which depends
only on the shape of the lattice; we write $\gamma = s/B$ for a general
squared harmonic $s = G^2$. The free energy per unit area is
\begin{equation}
F = -\epsilon\,\eta^2 + \Big[\tfrac12 K_{G=0} + C(\tau;\beta)\Big]\eta^4~,
\qquad
C(\tau;\beta) = \tfrac12\sum_{G\neq0} e^{-\gamma_G/2}\,K(\gamma_G B_c(\beta);\beta)~,
\end{equation}
where $\epsilon \propto B - B_c$ and $\tau$ labels the lattice shape
(section~\ref{sec:5.2}). The uniform term $K_{G=0}$ is the same for every
lattice, and the shell sum $C$ selects the lattice.

The shell sum is built from the quartic kernel $K(s;\beta)$, the self-interaction
of the condensate through a density modulation of wavenumber $G$. The
density sources the abelian field and the metric, and since the background
depends on $r$ only, the response at each $G$ is a radial boundary value
problem depending on $s$ alone (appendix~\ref{app:response}). The kernel is
\begin{equation}
K(s;\beta) = \big(C_4 - X\big)(s;\beta) - \tfrac12\,\mathcal{P}(s;\beta)~, \qquad
C_4 = \int p_1\,w_0^4\,dr~,
\end{equation}
with $p_1 = e^{Z-2V}$ and $p_2 = Ue^Z$ the two weights of the
Sturm--Liouville problem of section~\ref{sec:3}, $(p_2 w')' + Bp_1 w = 0$.
We normalise the zero mode on each brane by $\int p_1 w_0^2\,dr = 1$. The
contact term $C_4$ is the self-repulsion of the condensate and $X$ its
screening by the induced $U(1)$ field. In the probe limit this bracket is
the whole kernel, and its positivity is what makes the vortices repel and
form a lattice. The new term $\mathcal{P}$ is the exchange of the metric
response between two insertions of the condensate stress tensor. Its
weight $\tfrac12$ relative to $X$ is fixed by the quadratic on-shell action
(appendix~\ref{app:response-metric}). We write $X_{\rm ab}$ for the
screening by the photon alone with the metric frozen; it equals $X$ in the
probe limit (appendix~\ref{app:response-abelian}). Pairing the metric
response with the photon's Maxwell stress as well gives the dressed
pairing $\mathcal{P}_{\rm dress}$, and $K = C_4 - X_{\rm ab} -
\tfrac12\mathcal{P}_{\rm dress}$ splits the kernel into gauge-invariant
parts (appendix~\ref{app:response-metric}). On the grid
$0 < \beta \le 45$ described below and on the throat,
$\mathcal{P}_{\rm dress} > 0$: gravity lowers
the kernel below the photon-only kernel $C_4 - X_{\rm ab}$, as the
screening lowers it below $C_4$.

The uniform term $K_{G=0}$ is the response to a strictly uniform
condensate at fixed temperature and field. It is a separate radial problem,
in which the photon does not respond, since the Bianchi identity fixes the
uniform part of $F_{xy}$ at $B$. We computed it directly and find that it
equals the long-wavelength limit of the kernel,
\begin{equation}
K_{G=0}(\beta) = K_0(\beta) \equiv \lim_{s\to0}K(s;\beta)~,
\end{equation}
to $10^{-8}C_4$ on every rung of the ladder up to $\beta = 10^8$; we write $K_0$ for both from here on. The agreement is not
automatic, because the gauge used at finite $G$ becomes singular as
$G \to 0$; appendix~\ref{app:response-metric} shows that the pairing is
insensitive to that singular gauge transformation. The quartic coefficient of a uniform lattice is
therefore $\tfrac12K_0 + C$.

\begin{figure}[t]
\centering
\includegraphics{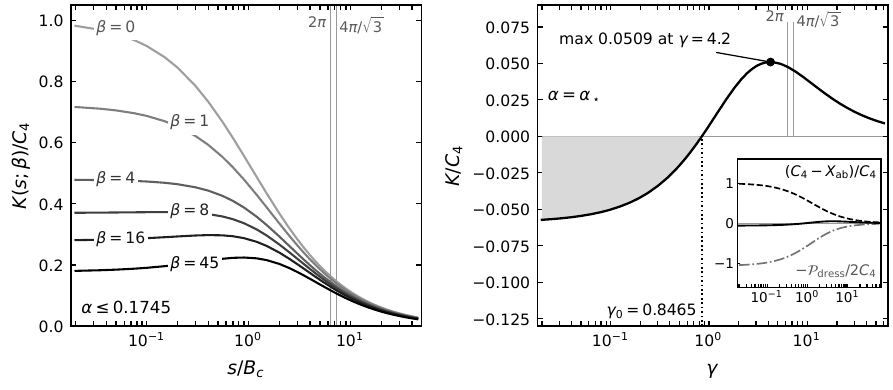}
\caption{The quartic kernel. Left: $K(s;\beta)/C_4$ against $s/B_c$ for
$\beta = 0, 1, 4, 8, 16, 45$ (light to dark), with the first shells $2\pi$
(square) and $4\pi/\sqrt3$ (triangular) marked. Right: the throat kernel
$K/C_4$ at $\astar$ against $\gamma = s/B_c$,
negative below $\gamma_0 = 0.8465$ (shaded), with (inset) its gauge-invariant
split (appendix~\ref{app:response-metric}): the photon pairing
$(C_4 - X_{\rm ab})/C_4$ with the metric frozen, and the dressed metric
pairing $-\mathcal{P}_{\rm dress}/2C_4$, whose near-cancellation
produces it.}
\label{fig:kernel}
\end{figure}

We computed the kernel on a dense grid up to $\beta = 45$ ($\alpha = 0.1745$), at selected harmonics along the ladder up to its rung
$\beta = 10^8$ ($\alpha = 0.544$), and at $\alpha_\star$
as a fixed problem on the throat (appendix~\ref{app:numerics-scan}). On the
grid it is positive everywhere but suppressed at long wavelength: $K/C_4$
at $s = 0.02B_c$ falls from $0.98$ at $\beta = 0$ to $0.18$ at $\beta = 45$,
and from $\beta \approx 8$ the kernel has a shallow interior maximum at
$s \lesssim B_c$ (figure~\ref{fig:kernel}). At $\alpha_\star$ it is negative
at long wavelength.

\subsection{The triangular lattice is the best Bravais lattice up to \texorpdfstring{$\alpha_I$}{alpha\_I}}
\label{sec:5.2}

Up to rotation and scale, a one-flux-quantum Bravais lattice is fixed by a
modular parameter $\tau = \tau_1 + i\tau_2$ in the upper half plane, with
reciprocal harmonics
\begin{equation}
\gamma_{mn} = \frac{2\pi}{\tau_2}\,|m + n\tau|^2~, \qquad (m,n) \neq (0,0)~.
\end{equation}
The set of harmonics is invariant under $\tau \to \tau + 1$ and
$\tau \to -1/\tau$, so $C$ is modular invariant for any kernel. The
triangular lattice $\tau = e^{i\pi/3}$ and the square lattice $\tau = i$,
the elliptic fixed points, are therefore stationary at every $\beta$.

In the probe limit the kernel is completely monotone in $s$, and
Montgomery's
theorem~\cite{Montgomery:1988mt,Cohn:2007universally,Betermin:2016theta}
makes the triangular lattice the global minimum over all one-flux-quantum
Bravais lattices (appendix~\ref{app:probe}). On the brane we know of no
Stieltjes representation for the metric pairing, so the proof of
appendix~\ref{app:probe} does not carry over. The function the theorem
needs, $e^{-\gamma/2}K(\gamma B_c)$, is in fact not completely monotone from
$\beta \approx 4$, where it turns concave at long wavelength, and from
$\beta \approx 8$ the kernel itself is not monotone. This does not by itself disfavour the triangular
lattice~\cite{Betermin:2016theta,Betermin:2019noncm}, which fully
backreacted holographic lattices in a magnetic field also
prefer~\cite{Donos:2015eew,Donos:2020viz}.
We scanned $C$ over the fundamental domain up to $\tau_2 = 5$ on every
brane of the grid and every rung of the ladder up to $\beta = 10^8$
($\alpha = 0.544$; appendix~\ref{app:numerics-scan}). On this grid the minimiser never leaves the triangular
point, and the Hessian there stays positive and
isotropic. We measure the selection by the relative margin over the square
lattice, $(C_\square - C_\triangle)/C_\triangle$, which falls slowly from
$0.246$ in the probe limit to $0.227$ at $\alpha = 0.1745$ and $0.18$ at
$\beta = 10^8$. The shape term itself is small,
$C_\triangle \approx 0.01\,C_4$, so in the probe limit this margin is a
$0.56\%$ difference in condensation energy.

Towards the stripe edge $C$ grows in proportion to $\sqrt{\tau_2}$ times a
Gaussian average $I(\beta)$ of the kernel over momentum transfers of order
$B_c$ (appendix~\ref{app:numerics-scan}). This average stays
positive up to $\alpha_I = 0.494$, so for $\alpha < \alpha_I$, $C$ grows
towards the stripe edge, and with the edges scanned to $\tau_2 = 50$
(appendix~\ref{app:numerics-scan}) the triangular minimum is global. Above $\alpha_I$,
$C$ is unbounded below towards the stripe edge, a sign that the truncation
has left its domain of validity rather than a prediction of infinite
elongation. Since $\alpha_I > \alpha_L$, this only concerns lattices that
are already unstable (section~\ref{sec:5.3}).

At $\alpha_\star$ the selection becomes a $\beta$-independent problem on the
BTZ throat, because the frozen flux plane holds $s/B$ fixed
(appendix~\ref{app:response-throat}). The triangular lattice
still beats the square, with relative margin $+0.1766$, but the kernel is negative for
$\gamma < \gamma_0 = 0.8465$, because the lighter of the two neutral
channels of the metric response has a longer range than the photon that
screens the repulsion. The brane kernel on the ladder converges onto these
values (appendix~\ref{app:numerics-scan}).

\subsection{The uniform lattice destabilises at \texorpdfstring{$\alpha_L = 0.372$}{alpha\_L = 0.372}}
\label{sec:5.3}

\begin{figure}[t]
\centering
\includegraphics{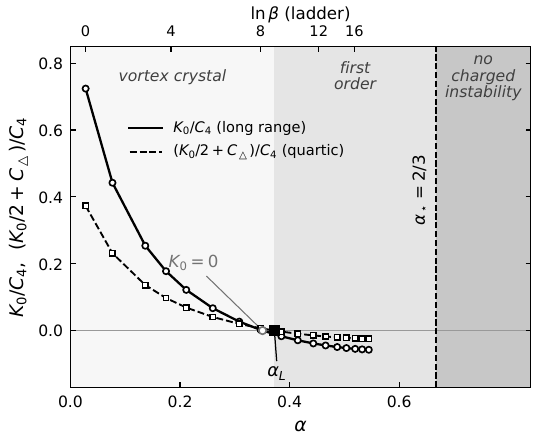}
\caption{The long-wavelength kernel along the ladder: $K_0/C_4$ (circles)
and the quartic coefficient of the uniform lattice
$(\tfrac12 K_0 + C_\triangle)/C_4$ (squares) on the rungs $\beta = 1$ to
$10^8$ against $\alpha$, with $\ln\beta$ of the ladder on the top axis. The
long-range interaction turns attractive where $K_0$ changes sign, at
$\alpha = 0.351$; the uniform lattice destabilises, and the transition
turns first order, where the quartic coefficient does, at
$\alpha_L = 0.372$. The crystal and first-order regimes, and the regime
without a charged instability, are shaded.}
\label{fig:density}
\end{figure}

Along the ladder $K_0/C_4$ falls monotonically and changes sign at
$\alpha = 0.3506$, $\beta = 3305$ (figure~\ref{fig:density}).
Beyond this coupling the interaction between density modulations of the
condensate is attractive at long wavelength. In the boundary theory this
is the exchange of the plasma's energy and pressure fluctuations between
the modulations. On the ladder $K_0/C_4$ saturates near $-0.06$, the value
of the throat kernel as $\gamma \to 0$ (figure~\ref{fig:kernel}, right).

A long-range attraction does not by itself destabilise the lattice, because
modulating the condensate density also modulates the lattice harmonics,
whose local energy resists it. Modulating the density of the uniform
lattice as $1 + \delta\cos(q\cdot x)$ on scales long compared with the
magnetic length adds harmonics at $G \pm q$ as well as at $q$, and to
second order the quartic energy rises by
\begin{equation}
\Delta F = \tfrac14\,\delta^2\eta^4 \sum_G e^{-\gamma_G/2}\,K\big(|G+q|^2;\beta\big)
\;\longrightarrow\; \tfrac14\,\delta^2\eta^4\,\big[K_0 + 2C_\triangle\big]
\qquad (q \to 0)~.
\end{equation}
The inverse compressibility of the lattice at fixed flux is thus
$K_0 + 2C_\triangle$, not $K_0$. To check that no other perturbation goes
first, we computed the second variation of the quartic functional about the
uniform lattice over all lowest-Landau-level perturbations at one flux
quantum. It splits over the magnetic Brillouin zone into $2\times2$ blocks
(appendix~\ref{app:response-hessian}). As $q \to 0$ the lower
branch is the Goldstone mode of translations and the upper branch tends to
$2(K_0 + 2C_\triangle)$. On a dense grid over the zone, from the probe
limit to $\alpha_L$, no eigenvalue is negative (in the probe limit the
lowest eigenvalue on the zone boundary is $0.0186\,C_4$, at the midpoint
$M$ of an edge). The
first instability of the uniform triangular lattice is therefore the
long-wavelength density mode, at $K_0 + 2C_\triangle = 0$, where the
quartic coefficient $\tfrac12K_0 + C_\triangle$ changes sign, that
is
\begin{equation}
\alpha_L = 0.3722
\end{equation}
(table~\ref{tab:couplings}).

Since $K_{G=0} = K_0$, the same combination is the quartic coefficient of
the uniform lattice, so $\alpha_L$ is also where the transition changes
order: it is a tricritical point of the onset line. It plays the role of
$\kappa = 1/\sqrt2$~\cite{Abrikosov:1956sx} in Ginzburg--Landau theory with a dynamical gauge
field, where the sign of the same Abrikosov quartic coefficient at
$H_{c2}$ separates type II from type I. Below $\alpha_L$ the normal phase gives way at $B_c$, in a
second-order transition, to a stable triangular lattice; between
$\alpha = 0.351$ and $\alpha_L$, where $K_0 < 0$, it is held stable by the
$2C_\triangle$ term. Above $\alpha_L$
the quartic coefficient is negative: the transition is first order, $B_c$
is the spinodal of the normal phase, and the uniform lattice is itself
unstable to long-wavelength modulation. The coefficient stays negative as
far as we computed the kernel, with
$(\tfrac12K_0 + C_\triangle)/C_4 = -0.025$ at $\beta = 10^8$, and what
replaces the lattice there is outside the quartic expansion. No other
one-flux-quantum Bravais lattice is stable above $\alpha_L$ either. Less
compact lattices keep their density stiffness $K_0 + 2C(\tau)$ longer; the
square keeps it to $\alpha = 0.3767$. But the square is a saddle of $C$ and
is unstable at the zone boundary at every coupling, and other shapes are
not stationary. Above $\alpha_L$, every lattice that still has positive
density stiffness is unstable to shear (appendix~\ref{app:numerics-scan}).\footnote{The electric
$p$-wave transition of~\cite{Ammon:2009xh} also turns first order beyond a
coupling, $\alpha \approx 0.13$ in our variable.}

\begin{table}[t]
\centering
\begin{tabular}{lccc>{\raggedright\arraybackslash}p{0.40\textwidth}}
\toprule
coupling & $\beta$ & $B_c u_H^2$ & $T_c/T_c^{\rm probe}$ & what changes there \\
\midrule
$0.351$ & $3.31\times10^3$ & $97.1$ & $0.23$ & long-wavelength vortex interaction turns attractive, $K_0 = 0$ \\
$\alpha_L = 0.372$ & $6.51\times10^3$ & $132.3$ & $0.20$ & uniform lattice unstable; transition first order, $B_c$ a spinodal \\
$\alpha_\star = 2/3$ & $\infty$ & $\infty$ & $0$ & no linear instability of the charged sector at any $B$, $T$ \\
\bottomrule
\end{tabular}
\caption{The couplings that organise the lattice and the onset, with the
backreaction parameter and the critical field at fixed temperature on the
onset curve, and the onset temperature at fixed field relative to its probe
value, $T_c/T_c^{\rm probe} = (5.13/B_c u_H^2)^{1/2}$. Below $\alpha_L$ a
triangular vortex crystal forms in a second-order transition; whether a
condensed phase exists above $\alpha_\star$ is open.}
\label{tab:couplings}
\end{table}

At fixed $\alpha$ the only scale is the temperature, so the onset is
$B_c = f(\alpha)\,T^2$ exactly, with $f = \pi^2 B_c u_H^2$ from
figure~\ref{fig:1}. Cooling at fixed field therefore meets it at
$T_c = \sqrt{B/f(\alpha)}$ for every $\alpha < \alpha_\star$. Gravity lowers
$T_c$ to a fifth of its probe value by $\alpha_L$
(table~\ref{tab:couplings}) and to under two percent of it at
$\alpha = 0.55$, and $T_c$ vanishes with the essential singularity of
section~\ref{sec:4.6} at $\alpha_\star$.
\section{Open questions}
\label{sec:6}

We leave three questions for future work.

The first, and the main one this work raises, is the phase diagram above
$\alpha_L$. There the quartic
coefficient of the uniform lattice is negative, so the transition is first
order and the onset $B_c$ is only the spinodal of the normal phase. Locating
the thermodynamic transition and the state it leads to needs the sextic
term on the brane, or the nonlinear condensed branch. In particular,
nothing we have computed bounds the first-order line to $\alpha < \alpha_\star$:
above $\alpha_\star$ the normal phase has no linear charged instability, but
a first-order transition into a condensed phase is not excluded, so
$\alpha_\star$ is established as the end of the instability and not yet as
the end of the vortex lattice.

The second is a proof of triangular minimality at finite coupling.
Numerically the triangular lattice is the global minimum over Bravais
lattices at every coupling we have solved below $\alpha_I$, but
Montgomery's theorem covers only the probe limit: on the brane the weighted
kernel is not completely monotone from $\beta \approx 4$, and below that no
proof is known (section~\ref{sec:5.2}). A proof would need a criterion for such kernels, such as
the sufficient condition of~\cite[theorem~1.1]{Betermin:2016theta}.

The third is the neutral sector. Above $\alpha_\star$ the charged sector
is mode stable: it has no growing quasinormal mode
(section~\ref{sec:4.5}). In this paper the fluctuations of $A^3$ and of
the metric about the brane have been analysed only in the static sector
sourced by the condensate (appendix~\ref{app:response-throat} and section~\ref{sec:5}), and
nonlinear stability is not addressed.

\acknowledgments

This work returns to an open question from the author's doctoral
research~\cite{Strydom:2014mfo},
carried out at the Max-Planck-Institut f\"ur Physik (Werner-Heisenberg-Institut),
F\"ohringer Ring 6, 80805 M\"unchen, Germany, for a doctorate from the
Ludwig-Maximilians-Universit\"at M\"unchen. The author thanks his former
collaborators Yan-Yan Bu, Johanna Erdmenger, Jonathan P. Shock, Stephan
Steinfurt and Hansj\"org Zeller.

The derivations, numerical code, symbolic checks, figures and text of this
paper were developed together with Claude, Anthropic's AI model (Claude
Fable 5, Fable 5.1 and Opus 5.5), which carried out the calculations, wrote
and ran the scripts that produce every number and figure, and drafted and
formatted the manuscript under the author's direction. Each result was
checked by independent methods: symbolic verification of the analytic
steps, and for each computed number a second, independent computation,
with either a different method or a different discretisation of the same
equations. The code
and its test suite are public. The author reviewed the derivations and the
text and checked the references, and takes responsibility for the whole.

\paragraph{Data and code availability.} The code that produces every
number and figure in this paper, the cached numerical data it reads, and a
test suite that checks the published values against the code are available
at \coderepo, and version~1.0.0 is archived on
Zenodo~\cite{Strydom:2026zenodo}. A single script there regenerates all
results from scratch.

\appendix
\section{The dictionary}
\label{app:dictionary}

\subsection{The D'Hoker--Kraus brane}
\label{app:dictionary-brane}

The background of section~\ref{sec:2.2} is the
uncharged magnetic brane of~\cite{DHoker:2009mmn,DHoker:2009ixq}. Their
action is $\frac{1}{16\pi G_5}\int\sqrt{-g}\,(R + 12 - F^2)$, so that their
Maxwell term corresponds to $\alpha = 2$ in our normalisation and their
Newton constant to $2\kappa^2 = 16\pi G_5$; their field $B$ is ours (the
$\mathcal{B} = \sqrt3 B$ of their boundary flux is a counting convention and
their Chern--Simons term vanishes identically for a constant field). We fixed
each factor by two independent routes, the Maxwell coefficient from their
Einstein equations and horizon constraints and the Newton constant from the
Schwarzschild free energy at $B = 0$, without reference to the observables
being compared. With this dictionary our fixed-temperature entropy shift
reproduces theirs, and their throat value $e^{2V} = B/\sqrt3$ is
\eqref{eq:throat} at $\alpha = 2$.

\subsection{The central charges}
\label{app:dictionary-charges}

In the conventions
of~\cite{Osborn:1993cr},
\begin{equation}
\langle T_{\mu\nu}(x)T_{\sigma\rho}(0)\rangle = \frac{C_T\,\mathcal{I}_{\mu\nu,\sigma\rho}}{x^{2d}}~, \qquad
\langle J_\mu(x)J_\nu(0)\rangle = \frac{C_J\,I_{\mu\nu}}{x^{2(d-1)}}~,
\end{equation}
with the current normalised by
the algebra of its charges, $[Q^a, Q^b] = i\epsilon^{abc}Q^c$, so that a doublet carries
$T^3 = \pm\tfrac12$. This is the boundary shadow of the bulk convention
$F^a = dA^a + \tfrac12\epsilon^{abc}A^b\wedge A^c$ of section~\ref{sec:2.1}. The two holographic normalisations quoted in
section~\ref{sec:4.4} are equation (3.15) of~\cite{Buchel:2009sk} at zero
Gauss--Bonnet coupling, with the action normalised as in their (2.1) and
their Planck length $\ell_P^{d-1} = \kappa^2$ and equations (30) and (54)
of~\cite{Freedman:1998tz}; the ratio $C_J/C_T = \tfrac{(d-1)(d-2)}{2d(d+1)}\alpha$
follows by division, and the map $\alpha \to \alpha\,(d-1)(d-2)/2d(d+1)$
turns $\alpha_\star(d) = 16/d(d-1)(d-2)$ into $8/d^2(d+1)$.

The check of section~\ref{sec:4.4} is the following. Romans' $SU(2)\times U(1)$
gauged supergravity~\cite{Romans:1985ps} is a consistent truncation around
the supergravity dual of any four-dimensional $\mathcal{N} = 2$
superconformal theory~\cite{Gauntlett:2007sm,Malek:2017njj}, and its $SU(2)$ gauge field
is dual to the $SU(2)_R$ current, whose two-point function is fixed by $c$
alone. In any four-dimensional conformal field theory
$C_T = 40c/\pi^4$~\cite{Osborn:1993cr},
and the $\mathcal{N} = 1$ superconformal Ward identity gives, for the
$\mathcal{N} = 1$ R-current, $C_R = 4c/\pi^4$, the supercurrent coefficient of
equations (3.43) and (11.12) of~\cite{Osborn:1998qu}, whose lowest component is the
R-current~\cite{Anselmi:1997am}. We decompose the $\mathcal{N} = 2$
R-symmetry as $R_{\mathcal{N}=1} = \tfrac13 r + \tfrac43 I_3$. The
combination $r - 2I_3$, under which one supercharge is neutral, sits in a
linear multiplet and is therefore orthogonal to the R-current. It follows
that $C_r = 8\,C_{I_3}$ for the $U(1)_r$ and $SU(2)_R$
Cartan currents, and hence
\begin{equation}
C_J(SU(2)_R) = C_{I_3} = \frac{3c}{2\pi^4}~, \qquad \frac{C_J(SU(2)_R)}{C_T} = \frac{3}{80}~,
\end{equation}
in every $\mathcal{N} = 2$ superconformal theory.

On the gravity side, the coefficient of the $SU(2)$ kinetic term in Romans'
Lagrangian, in the form of~\cite{Lu:1999bw} with the scalar at its
extremum $\mathcal{X} = 1$ and the gauge field rescaled to unit structure constant, is
$\alpha_{\rm SUSY} = L^2/4$; we obtained the same value from the
eleven-dimensional reduction of~\cite{Gauntlett:2007sm} and from Romans'
original conventions via~\cite{Ohl:2010au}. Then
$(3/20)(1/4) = 3/80$, which fixes both normalisations in $d = 4$ from the
field theory side. The scalar $\mathcal{X}$ is sourced by the $SU(2)$ field strength
at the same order as the metric, $(\Box + 4)\,\delta\mathcal{X} = -\tfrac1{24}F^aF^a$
at the linearised level in our normalisation, so $\mathcal{X}$ has
$m^2 = -4$ and $\Delta = 2$, which is why Einstein--$SU(2)$ theory with $SU(2)$
flux is not a truncation of Romans' theory.

The $d = 3$ check is the same construction one dimension down. On
the gravity side, minimal $\mathcal{N} = 2$ gauged supergravity in four
dimensions has the geometrised Maxwell normalisation $-F^2$ relative to
$R$ and a gravitino of charge $1/\ell$ on that field~\cite{Caldarelli:1998hg};
the field of unit supercurrent charge is $A/\ell$, so
$\alpha_{\rm SUSY} = 2\ell^2 = 2$. The reduction of eleven-dimensional
supergravity on any Sasaki--Einstein seven-manifold~\cite{Gauntlett:2007ma}
gives $\alpha = 1/2$ for its gauge field, under which the gravitino has
charge $1/2$ (by matching the action to~\cite{Caldarelli:1998hg}), and rescaling to unit charge gives $2$ again. The charge
normalisation is fixed independently by requiring that the
Bogomol'nyi--Prasad--Sommerfield Reissner--Nordström solution saturate
$\Delta = R$, the chiral-primary bound of the boundary theory. On the field
theory side, the three-dimensional $\mathcal{N} = 2$ superconformal Ward
identity writes both the $R$-current and the stress-tensor two-point
functions in terms of a single coefficient~\cite{Closset:2012ru},
which in the conventions of~\cite{Osborn:1993cr} gives $C_R/C_T = 1/6$ for
every such theory, and a free chiral multiplet with $R$-charges
$(\tfrac12, -\tfrac12)$ reproduces it by explicit Wick contraction,
$C_R = 1/16\pi^2$ and $C_T = 3/8\pi^2$. The holographic formulas at
$d = 3$ give $C_J/C_T = \alpha/12$, and $2/12 = 1/6$.
\section{The reduction of the charged sector}
\label{app:charged}

\subsection{The zero mode in general dimension}
\label{app:charged-zero-mode}

On the diagonal brane
$ds^2 = -U dt^2 + dr^2/U + e^{2V}(dx^2 + dy^2) + e^{2Z}d\vec z^{\,2}$ in
$d + 1$ dimensions, with $d - 3$ directions $\vec z$ along the field, the
Ricci-form equations are $R_{MN} + d\,g_{MN} = \alpha(F^a_{MP}F^{a\phantom{N}P}_N
- F^2 g_{MN}/2(d-1))$. We write $W = A^1 + iA^2$ and $D = \partial + iA^3$,
so that on $y$-independent functions the Landau ladder is
$D_\pm = D_x \pm iD_y = \partial_x \mp Bx$ with
$D_+\psi_n = -\sqrt B\,\psi_{n+1}$, $D_-\psi_n = 2n\sqrt B\,\psi_{n-1}$ and
$\psi_n = H_n(\sqrt B x)\,e^{-Bx^2/2}$. For the polarised lowest Landau
level $W_+ \equiv W_x + iW_y = p_+(r)\psi_0$, $W_- \equiv W_x - iW_y = 0$, $W_r = 0$, the
linearised Yang--Mills equations reduce, after the $x$ integration is done
exactly, to
\begin{equation}
\big(P w'\big)' + Q w = 0~, \qquad P = U e^{(d-3)Z}~, \qquad Q = B\,e^{(d-3)Z - 2V}~,
\end{equation}
with $w = p_+/2$. The transverse Landau term vanishes because
$F^W_{xy} = \tfrac{i}{2}\big(D_+ W_- - D_- W_+\big)$ has no component in this polarisation, and
$Q$ is the non-abelian moment term $\epsilon^{abc}F^b_{MN}W^{cM}$ with the
single warp factor $e^{-2V}$ of the flux plane. At the same order nothing
sources $A^3$ or the metric, which we verified term by
term, so the truncation is consistent. Near the boundary the two solutions
behave as $w = j_0\big(1 + \tfrac{B}{2}r^{-2}\ln r + \ldots\big) + j_1 r^{-2}
+ \ldots$ in $d = 4$, and the source-free condition is $j_0 = 0$. The
brane's own corrections start at $r^{-3}$ in the fixed-temperature frame,
where $U = r^2 + 2ar + \ldots$, so they are subleading to both terms, and
the source can be read off unambiguously as long as the fit or the
extraction allows for them. At the
horizon, regularity of the solution with $P \sim U'(r_p)(r - r_p)$ requires
$w'(r_p) = -Q(r_p)\,w(r_p)/P'(r_p)$.

\subsection{The other channels}
\label{app:charged-channels}

For the argument of section~\ref{sec:4.5} we keep the
general static, $z$-independent fluctuation in the gauge $W_r = 0$. Since
$D_\pm$ shift the Landau index by one and the background depends on $r$
only, the fluctuations sort by a single integer $n \ge 0$,
\begin{equation}
W_+ = p_+(r)\,\psi_n~, \quad W_- = p_-(r)\,\psi_{n-2}~, \quad W_t = a(r)\,\psi_{n-1}~, \quad W_z = c(r)\,\psi_{n-1}~,
\end{equation}
with $\psi_{k<0} \equiv 0$. Integrating the quadratic action over $x$ is
exact at every $n$ and produces no coupling between different $n$. The
components $W_t$ and $W_z$ decouple from $(p_+, p_-)$ and from each other, each
with a Sturm--Liouville operator $(P_c a')' - (2n-1)B\zeta_c\,a$ whose
coefficients $P_c, \zeta_c$ are positive and whose transverse mass
$(2n-1)B$ is positive for $n \ge 1$. In the convention of appendix~\ref{app:charged-zero-mode}, with
$P = \sqrt{-g_{tt}g_{zz}/g_{rr}}$, $\zeta = \sqrt{-g_{tt}g_{rr}g_{zz}}/g_{xx}$
(on the brane $P = p_2$ and $\zeta = p_1 = Q/B$, the weights of
section~\ref{sec:5}) and $N_k = \int\psi_k^2\,dx$, the remaining density, written as minus the
energy so that a positive $T_n$ is destabilising, is
\begin{equation}
\mathcal{E}_n = -\frac{P}{4}\big(N_n\,p_+'^2 + N_{n-2}\,p_-'^2\big) + T_n(p_+, p_-)~, \qquad
T_0 = +\frac{B\zeta N_0}{4}\,p_+^2~, \qquad T_1 = 0~,
\end{equation}
and, for $n \ge 2$,
\begin{equation}
\label{eq:Tn}
T_{n} = -\frac{B\zeta N_{n-1}}{4}\begin{pmatrix} 2n(n-1) & n \\ n & \dfrac{n}{2(n-1)} \end{pmatrix}~,
\end{equation}
which has vanishing determinant and negative trace. There are no cross
terms between derivatives and undifferentiated fields, and the real and
imaginary parts of $(p_+, p_-)$ carry identical blocks. The null vector of
$T_{n\ge2}$, $(p_+, p_-) \propto (-1, 2(n-1))$, is the residual gauge
transformation $W_M = D_M(\theta\psi_{n-1})$, as the ladder relations
$D_+\psi_{n-1} = -\sqrt B\psi_n$ and $D_-\psi_{n-1} = 2(n-1)\sqrt B\psi_{n-2}$
show, and with $W_r$ reinstated the density shifts by a total derivative
under it, which we used as the check on the reduction. For general $n$ the transverse form follows from the ladder relations
alone: the transverse energy is $\tfrac12\zeta$ times
$\tfrac14|D_+W_- - D_-W_+|^2 - \tfrac B2\big(|W_+|^2 - |W_-|^2\big)$, and with $D_+\psi_{n-2} = -\sqrt B\psi_{n-1}$,
$D_-\psi_n = 2n\sqrt B\psi_{n-1}$ and $N_n = 2nN_{n-1}$ the $x$ integral
gives $\tfrac{B\zeta N_{n-1}}{4}\big[2n(n-1)p_+^2 + 2np_+p_- + \tfrac{n}{2(n-1)}p_-^2\big]$,
which is $-T_n$ of~\eqref{eq:Tn}, as a symbolic reduction for
$n = 0, \ldots, 4$ on the general diagonal metric confirms.
The statement of section~\ref{sec:4.5} follows: for $n \ge 1$ the energy is a sum of
positive radial terms and a positive semi-definite transverse form, and it
vanishes only on $r$-independent pure gauge, so a source-free, horizon-regular
zero mode must be pure gauge, while $n = 0$, with $p_+ = 2w$, reproduces
$N_0$ times the density $-Pw'^2 + Qw^2$, whose Euler--Lagrange equation
is the zero-mode equation of appendix~\ref{app:charged-zero-mode}.

The argument of section~\ref{sec:4.5} is made for static, $z$-independent
perturbations. With a dependence $e^{ik_zz}$ the $(x,z)$ and $(y,z)$ terms
of the energy become
$\tfrac14\sqrt{-g}\,g^{xx}g^{zz}\big(|D_+W_z - ik_zW_+|^2 + |D_-W_z - ik_zW_-|^2\big)$,
two squares that mix $W_z$ with $W_\pm$, while the $(x,y)$ form is unchanged.
For $n \ge 1$ the energy therefore stays non-negative, with kernel the pure
gauge $D_M(\theta\psi_{n-1}e^{ik_zz})$, and for $n = 0$, where
$W_- = W_z = 0$, the new term is the positive mass $N_0 e^{-Z}k_z^2w^2$.
A dependence on $z$ cannot produce a zero mode where the $k_z = 0$ operator
has none.

The onset is also static. The background has no $A_t$ and no $g_{ti}$, so
time derivatives enter the quadratic action only through $|F_{ti}|^2$:
\begin{equation}
L = \tfrac12\sqrt{-g}\,(-g^{tt})g^{ii}|F_{ti}|^2 - \mathcal V[W]~,
\end{equation}
where $\mathcal V$ is the static energy above. The magnetic field makes the
operator of $\mathcal V$ Hermitian rather than real symmetric, but it does
not add a term linear in $\partial_t$. Take a mode $e^{-i\omega t}$ with
$\mathrm{Im}\,\omega > 0$, ingoing and source-free, in the gauge $W_t = 0$.
The components of $W$ in this gauge then vanish at the horizon like
$(r - r_p)^{\mathrm{Im}\,\omega/U'(r_p)}$; $F_{ri}$ grows like
$(r - r_p)^{\mathrm{Im}\,\omega/U'(r_p) - 1}$, but every integrand goes
like $(r - r_p)^{2\mathrm{Im}\,\omega/U'(r_p) - 1}$ and every surface term
like $(r - r_p)^{2\mathrm{Im}\,\omega/U'(r_p)}$. Contracting the equations
with $\bar W$ therefore gives, without surface terms,
\begin{equation}
\frac{\omega^2}{|\omega|^2}\int\sqrt{-g}\,(-g^{tt})g^{ii}|F_{ti}|^2 = 2\,\mathcal V[W]~.
\end{equation}
Hence $\omega^2$ is real, so $\omega = i\Gamma$, and the static energy of
the mode is negative. For $n \ge 1$, $\mathcal V \ge 0$ and no mode grows,
also at $k_z \neq 0$: $W_r$, which this gauge leaves in place, enters
$\mathcal V$ only through the squares $|F_{ri}|^2$, and the transverse form
does not contain it. For $n = 0$ the mode obeys
$(Pw')' + Qw = \Gamma^2 R\,w$ with $R = e^{Z}/U$ positive,
a self-adjoint problem whose continuum begins at $\Gamma^2 = 0$; by Sturm
oscillation the number of its growing modes equals $\#\{k : B_k < B\}$,
and they emerge through $\omega = 0$. A chemical potential would add a term linear in
$\partial_t$ and shift $\omega$ to $\omega - \mu$, which is where the
argument stops.

We checked the $n = 0$, $k_z = 0$ channel numerically on the brane at
$\beta = 1$, $45$ and $10^4$ and in the probe limit, by collocation in
ingoing coordinates and independently by the argument principle applied to
the shooting source coefficient. Both count exactly
$\#\{k : B_k < B\}$ modes in the upper half plane; the collocation
eigenvalues are purely imaginary to
$2\times10^{-11}$, and one of them crosses $\omega = 0$ exactly at $B_c$.
\section{The response system and the quartic kernel}
\label{app:response}

\subsection{The condensate as a source}
\label{app:response-source}

Just above the onset, $W_x = \eta\,
w_0(r)\,\psi(x,y)$ with $W_y = -iW_x$, and $|\psi|^2 = \sum_G \omega_G\,
e^{iG\cdot x}$ with $\omega_G = (-1)^{m+n+mn}e^{-\gamma_G/4}$ on the reciprocal
lattice, $G = G_{mn}$, so that the quartic free energy carries the weights
$|\omega_G|^2 = e^{-\gamma_G/2}$.
At second order in $\eta$ the condensate sources the abelian field and
the metric at each $G$, through the current $\epsilon^{3bc}W^b\,DW^c$ and the
stress tensor of $W$ evaluated on the zero mode. Both close on
$|\psi|^2$ alone, so the response to the harmonic $\eta^2\omega_G\,e^{iGx}$, which we take along $x$, is
\begin{equation}
\begin{split}
h_{MN}\,dx^Mdx^N &= \eta^2\omega_G\,e^{iGx}\Big[-U H_{tt}\,dt^2 + e^{2V}\big(H_{xx}\,dx^2 + H_{yy}\,dy^2\big) + e^{2Z}H_{zz}\,dz^2 \\
&\qquad\qquad\qquad + \frac{H_{rr}}{U}\,dr^2 + 2ie^{2V}H_{xr}\,dx\,dr\Big]~, \\
\delta A^3 &= \eta^2\omega_G\,e^{iGx}\,a_y\,dy~, \qquad \hat b = iG\,a_y~,
\end{split}
\end{equation}
with $\hat b$ the induced field strength. Per unit $\eta^2\omega_G e^{iGx}$
the stress tensor of $W$, in the normalisation of the Einstein equations
$G_{MN} - 6g_{MN} = \alpha\,(F_{MP}F_N{}^P - \tfrac14 g_{MN}F^2)$, is diagonal,
\begin{equation}
\begin{split}
&S_{tt} = Ue^{-4V}\big(Ue^{2V}w_0'^2 - Bw_0^2\big)~, \qquad
S_{xx} = S_{yy} = -Be^{-2V}w_0^2~, \qquad \\
&S_{zz} = e^{2Z-4V}\big(Bw_0^2 - Ue^{2V}w_0'^2\big)~, \qquad
S_{rr} = \frac{e^{-4V}}{U}\big(Bw_0^2 + Ue^{2V}w_0'^2\big)~,
\end{split}
\end{equation}
and it is conserved on the zero-mode equation, $\nabla^M S_{MN} = 0$ up to
the Lorentz force on the condensate current, whose source for $\hat b$ is
$s\,p_1w_0^2$, $s = G^2$. Since the background depends on $r$ only, the response at
wavenumber $G$ is a radial boundary value problem in which $G$ enters
through $s$ alone, and the quartic free energy is the on-shell action of the
response paired against its own source.

\subsection{The abelian sector}
\label{app:response-abelian}

With $p_1 = e^{Z-2V}$ and $p_2 = Ue^Z$, the
induced abelian field $\hat b$ at harmonic $s$ obeys, with the metric frozen,
\begin{equation}
\big(p_2\,\hat b'\big)' = s\,p_1\big(\hat b - w_0^2\big)~,
\end{equation}
regular at the horizon and vanishing at the boundary, and
$X_{\rm ab}(s) = \int p_1 w_0^2\,\hat b\,dr$. Writing $\mathcal{A} = -\tfrac{d}{dr}p_2\tfrac{d}{dr}
\ge 0$, $\mathcal{M} = p_1$, $\mathcal{T} = \mathcal{M}^{-1/2}\mathcal{A}\mathcal{M}^{-1/2}$ and $\mathbf{u} = \mathcal{M}^{1/2}w_0^2$, this is
$(C_4 - X_{\rm ab})(s) = \langle \mathbf{u}, \mathcal{T}(\mathcal{T}+s)^{-1}\mathbf{u}\rangle$, a Stieltjes transform of a
positive measure, hence completely monotone in $s$ on every brane. The
identity $\tfrac{d}{dr}\big[p_2 w^3 w'\big] = 3p_2 w^2 w'^2 - Bp_1 w^4$ on
solutions of the zero-mode equation, whose endpoint terms vanish, gives
$\langle \mathbf{u}, \mathcal{T}\mathbf{u}\rangle = \int p_2\big((w_0^2)'\big)^2 dr = \tfrac43 B_c C_4$
and hence the tail $(C_4 - X_{\rm ab})(s) = \tfrac43 B_c C_4/s + O(s^{-2})$
on every brane.

\subsection{The coupled metric and photon response}
\label{app:response-metric}

The classifying symmetry of the
linearised system is $y$-parity combined with the flip of the $U(1)$ charge,
since the background $F_{xy}$ is odd under $y$-parity alone. Under it $a_y$
is even and $a_x$, $a_r$ are odd and unsourced. The diffeomorphisms
$\xi = (\xi^x, \xi^r)\,e^{iGx}$ of the even sector shift $H_{tt}$ by a
multiple of $\xi^r$ and $H_{xx}$ by multiples of $\xi^x$ and $\xi^r$, with
no derivatives, so the algebraic gauge $H_{tt} = H_{xx} = 0$ can be reached
pointwise. The linearised equations are
$\delta(G_{MN} - 6g_{MN}) - \alpha\,\delta T_{MN} = \alpha S_{MN}$, with
$\delta T$ the variation of the abelian stress in both the metric and the
field, and the $y$ component of the linearised Maxwell equation sourced by
$J$. Using the background constraint
$U'(2V' + Z') + 2UV'(V' + 2Z') = 12 - \alpha B^2e^{-4V}$ to simplify the
coefficients, the $(rr)$ and $(xr)$ components are first order,
\begin{equation}
\begin{split}
&\big(\tfrac12\alpha B^2e^{-4V} - 6\big)H_{rr} + \tfrac12 G\,(U' + 2UV' + 2UZ')\,H_{xr}
+ \tfrac14(U' + 2UV' + 2UZ')\,H_{yy}' \\
&\quad + \tfrac14(U' + 4UV')\,H_{zz}' - \tfrac12\big(\alpha B^2e^{-4V} + G^2e^{-2V}\big)H_{yy} - \tfrac12 G^2e^{-2V}H_{zz} \\
&\quad + \alpha Be^{-4V}\,\hat b = \alpha e^{-4V}\big(Bw_0^2 + Ue^{2V}w_0'^2\big)~,
\end{split}
\end{equation}
\begin{equation}
\begin{split}
&\tfrac14 G^2(U' + 2UV' + 2UZ')\,H_{rr} + \alpha B^2 G\,Ue^{-2V}H_{xr}
- \tfrac12 G^2U\big(H_{yy}' + H_{zz}'\big) \\
&\quad + \tfrac12 G^2U(V' - Z')\,H_{zz} + \alpha BUe^{-2V}\,\hat b' = 0~,
\end{split}
\end{equation}
the $(yy)$ and $(zz)$ components are second order,
\begin{equation}
\begin{split}
&\tfrac12 U H_{zz}'' + \tfrac12(U' + UV' + 2UZ')\,H_{zz}' - \tfrac12 G^2e^{-2V}H_{zz}
+ \tfrac12\alpha B^2e^{-4V}H_{yy} + GUH_{xr}' \\
&\quad + G(U' + 2UV' + UZ')\,H_{xr} - \tfrac14(U' + 2UV' + 2UZ')\,H_{rr}' \\
&\quad - \tfrac12\big(G^2e^{-2V} + \alpha B^2e^{-4V} + 12\big)H_{rr}
- \alpha Be^{-4V}\,\hat b = -\alpha Be^{-4V}w_0^2~,
\end{split}
\end{equation}
\begin{equation}
\begin{split}
&\tfrac12 U H_{yy}'' + \tfrac12(U' + 3UV')\,H_{yy}' - \tfrac12\big(\alpha B^2e^{-4V} + G^2e^{-2V}\big)H_{yy}
+ GUH_{xr}' \\
&\quad + G(U' + 3UV')\,H_{xr} - \tfrac14(U' + 4UV')\,H_{rr}' + \tfrac12\big(\alpha B^2e^{-4V} - G^2e^{-2V} - 12\big)H_{rr} \\
&\quad + \alpha Be^{-4V}\,\hat b = \alpha e^{-4V}\big(Bw_0^2 - Ue^{2V}w_0'^2\big)~,
\end{split}
\end{equation}
and the Maxwell equation is
\begin{equation}
\big(p_2\,\hat b'\big)' - s\,p_1\big(\hat b - w_0^2\big) + BG\big(p_2 H_{xr}\big)'
+ \tfrac12 sBp_1\big(H_{yy} - H_{zz} - H_{rr}\big) = 0~,
\end{equation}
which with the metric perturbation switched off is the abelian equation of appendix~\ref{app:response-abelian}.
These five rows, with the derivatives of the two first-order ones, form a
linear system for $(H_{rr}, H_{xr}, H_{rr}', H_{xr}', H_{yy}'', H_{zz}'', \hat b'')$
that is non-singular on the brane. The even sector therefore closes on the
three evolution fields $(H_{yy}, H_{zz}, \hat b)$, with $H_{rr}$ and $H_{xr}$
fixed algebraically; the $(tt)$ and $(xx)$ equations then vanish identically
on the solved forms, as the linearised Bianchi identity and source
conservation require. Regularity at the horizon fixes the slope of
each evolution field in terms of its value, and the regular solutions have
$H_{rr}(r_p) = 0$, so that $h_{rr} = H_{rr}/U$ is regular and the
temperature is not modulated. The boundary indicial exponents are
$\{0, 4\}$ for the metric fields and $\{0, 2\}$ for $\hat b$; the
normalisable solution is the one with the three leading coefficients zero,
and then $H_{yy}, H_{zz}, H_{rr} = O(r^{-4})$, $H_{xr} = O(r^{-5})$ and
$\hat b = O(r^{-2})$. At $\beta = 0$ the $(H_{yy}, H_{zz})$ block reduces to
the probe graviton system and the Maxwell row to the abelian equation.

\paragraph{The pairing.} Let $\mathcal{L}_q$ be the part of
$\sqrt{-g}\,\big(R + 12 - \tfrac{\alpha}{2}F^2\big)$ quadratic in the
response $(h, a_y)$ (the harmonic times its conjugate, averaged over
the cell), and add the couplings to the condensate,
\begin{equation}
\mathcal{L}_2 = \mathcal{L}_q + \alpha\sqrt{-g}\,S^{MN}\big(h_{MN} + \bar h_{MN}\big)
+ 2\alpha\,p_1w_0^2\big(\hat b + \bar{\hat b}\big)~,
\end{equation}
with indices raised by the background metric and $h_{MN}$ stripped of
$\eta^2\omega_Ge^{iGx}$. The coefficients of the two source terms are not
chosen: they are the unique ones for which the Euler--Lagrange equations of
$\mathcal{L}_2$ are the five rows above, with a single overall factor
$-\sqrt{-g}$ on the Einstein rows and $-2\alpha$ on the Maxwell row.
Since $\mathcal{L}_q$ is a homogeneous quadratic form, Euler's theorem gives
$\sum_\phi \phi\,\mathrm{EL}_\phi(\mathcal{L}_q) = 2\mathcal{L}_q - \mathcal{Q}'$, the sum running over the fields and their conjugates, with
\begin{equation}
\mathcal{Q} = \sum_\phi\Big[\phi\Big(\frac{\partial\mathcal{L}_q}{\partial\phi'}
- \Big(\frac{\partial\mathcal{L}_q}{\partial\phi''}\Big)'\Big)
+ \phi'\,\frac{\partial\mathcal{L}_q}{\partial\phi''}\Big]~,
\end{equation}
so that on shell $\mathcal{L}_2$ is half its source terms plus
$\tfrac12\mathcal{Q}'$. The flux vanishes at both ends. At the horizon every
term of $\mathcal{Q}$ carries a factor $U$ except
\begin{equation}
\begin{split}
\mathcal{Q}(r_p) = -\tfrac32\,U'e^{2V+Z}\Big[&H_{rr}\bar H_{rr} - \tfrac16 H_{rr}\big(\bar H_{yy} + \bar H_{zz}\big) \\
&- \tfrac16\bar H_{rr}\big(H_{yy} + H_{zz}\big)\Big]~,
\end{split}
\end{equation}
which vanishes with $H_{rr}(r_p)$; at the boundary it falls off as $r^{-2}$,
set by the photon. With the fall-offs above every local boundary term
quadratic in the response, the Gibbons--Hawking term and the holographic
counterterms including the logarithmic $F^2$ one, vanishes as the cutoff is
removed, and the horizon is not a boundary, so the quartic free energy is
the bulk on-shell action alone. In the units of the probe kernel,
where the $\beta \to 0$ limit is $C_4 - X_{\rm ab}$, the on-shell action is
multiplied by $-1/2\alpha$, and the source terms give
\begin{equation}
K(s;\beta) = C_4 - X(s;\beta) - \tfrac12\,\mathcal{P}(s;\beta)~,
\end{equation}
with
\begin{equation}
X = \mathrm{Re}\!\int p_1w_0^2\,\hat b\,dr~, \qquad
\mathcal{P} = \mathrm{Re}\!\int\sqrt{-g}\;S^{MN}\,\bar h_{MN}\,dr~,
\end{equation}
where $\hat b$ and $h$ are the coupled solution, including the algebraic
components, and the contact term $C_4$ is the quartic self-interaction of
$W$. The weight $\tfrac12$ on $\mathcal{P}$ relative to $X$ is the
ratio of the two source couplings in $\mathcal{L}_2$, which the field
equations fix. The difference between $X$ and the abelian-only $X_{\rm ab}$
of appendix~\ref{app:response-abelian} is the part of the photon response
induced through the metric, and it is $X_{\rm ab}$ that carries the spectral
representation and the tail identity.

\paragraph{The weight $\tfrac12$.} The weight has the familiar origin of
the matter couplings: a stress tensor couples to the metric through
$\tfrac12\sqrt{-g}\,T^{MN}h_{MN}$, with
$T^{MN} = (2/\sqrt{-g})\,\delta S/\delta g_{MN} = 2\alpha S^{MN}$ in the
normalisation above, while a current couples to the gauge field through
$\sqrt{-g}\,J^Ma_M$ with no such factor. Gauge invariance fixes it
independently of the action. Under $\xi = (\xi^x, \xi^r)\,e^{iGx}$ the response
shifts by $h \to h + \mathcal{L}_\xi g$ and $a_y \to a_y + B\xi^x$. For a
weight $w$ on $\mathcal{P}$, the shift of $X + w\mathcal{P}$ by $\xi^x$ is
proportional to $w - \tfrac12$ pointwise, because the $(xx)$ component of
the condensate stress is the magnetic pressure of the current that couples
to $a_y$. At $w = \tfrac12$ the whole shift is a total derivative,
\begin{equation}
\delta_\xi\Big(X + \tfrac12\mathcal{P}\Big) = \Big[\mathcal{Q}_\xi\Big]_{r_p}^{\infty}~, \qquad
\mathcal{Q}_\xi = e^{Z-2V}\big(Bw_0^2 + Ue^{2V}w_0'^2\big)\,\mathrm{Re}\,\xi^r~,
\end{equation}
which vanishes for every $\xi$ that keeps the horizon in place,
$\xi^r(r_p) = 0$ as fixed temperature requires, and grows more slowly than
$r^3$ at the boundary, which includes the boundary Weyl rescalings. The
kernel is therefore the same in every such gauge. Transforming the solutions to
radial gauge, $H_{rr} = H_{xr} = 0$, and to a generic smooth gauge confirms
this (table~\ref{tab:checks}).

Pairing the metric response instead with the full stress, including the
Maxwell stress $T_{MN}[\hat b_{\rm ab}]$ of the decoupled photon, gives the
dressed pairing $\mathcal{P}_{\rm dress}$, with
$X - X_{\rm ab} = \tfrac12\big(\mathcal{P}_{\rm dress} - \mathcal{P}\big)$,
because the cross term of $\mathcal{L}_q$ between $h$ and $a_y$ is $\alpha$
times the dressing pairing up to a total derivative. Hence
$K = C_4 - X_{\rm ab} - \tfrac12\mathcal{P}_{\rm dress}$, a split into
gauge-invariant parts, which agrees with $K$ only at weight $\tfrac12$.

The uniform response $K_{G=0}$ of section~\ref{sec:5.1} comes from a
separately derived $G = 0$ system in the gauge $H_{tt} = 0$, as a
solvability condition: it is the contact term plus the shift of the onset
eigenvalue by the condensate's own uniform metric response,
$K_{G=0} = C_4 + \delta B[h_0]$, and the weight $\tfrac12$ there follows
from the first variation of the coefficients of the zero-mode equation,
with no action identity. A weight $w$ on $\mathcal{P}$ would move $K_0$ by
$-1.105\,(w - \tfrac12)\,C_4$ at $\beta = 45$, so the agreement
$K_{G=0} = K_0$ fixes $w = \tfrac12$ to $10^{-8}$. That
agreement is not automatic: at small $G$ the gauge $H_{tt} = H_{xx} = 0$
is reached by a diffeomorphism along $x$ of size $1/G$, which drags the
photon with it and leaves a finite pure-gauge induced field as $s \to 0$.
By the gauge invariance above the pairing does not see it, and since the
boundary metric is held fixed at every $G$, the limit lands on the
fixed-temperature solution.

\subsection{The throat limit}
\label{app:response-throat}

In the fixed-temperature frame, $\beta \to \infty$
takes the interior to $U \to (r - 1)(3r + 1)$, $e^{2Z} \propto (r - \tfrac13)^2$
and $e^{2V} \to \sqrt{\beta/6}$, that is BTZ with $h = 2/3$, $r_0 = 1/3$. With $\rho = r - r_0$, every ingredient of the kernel is governed
by the single radial operator
\begin{equation}
L_m\,y = \big[\rho(\rho^2 - h^2)\,y'\big]' - m\,\rho\,y~,
\end{equation}
which in $\varsigma = h^2/\rho^2$ becomes
$4\varsigma^2(1-\varsigma)y'' - 4\varsigma^2y' - my = 0$, with exponents
$\{0, 0\}$ at the horizon and $(1 \mp \mu)/2$ at the top, $\mu = \sqrt{1+m}$,
and the horizon-regular and normalisable solutions are
$y_H = \varsigma^a\,{}_2F_1(a, a; 1; 1 - \varsigma)$ and $y_I = \varsigma^a\,{}_2F_1(a, a; 1 + \mu; \varsigma)$,
with $a = (1+\mu)/2$. The zero mode $L_{-1}w_0 = 0$ is $m = -1$, $\mu = 0$,
for which the horizon-regular solution is
$w_0 \propto \tfrac{2}{\pi}\sqrt{\varsigma}\,\mathbf{K}(1 - \varsigma)$, with
$\mathbf{K}$ the complete elliptic integral of the first kind; its amplitude
is not normalisable in the throat and cancels from every lattice comparison,
while the quartic measure $p_1 w_0^4\,dr$ converges at the top of the throat.
At finite $\beta$ the same formula
with $a = (1 + i\nu)/2$ supplies the horizon phase $\chi(\nu)$ of section~\ref{sec:4.6}.
The induced field at harmonic $s$ obeys $L_\lambda\hat b = -\lambda\rho
w_0^2$ with $\lambda = s\ell_3^2 e^{-2V} = \gamma b$. The abelian throat
kernel $\mathcal{S}(\gamma) = C_4 - X_{\rm ab}$ is taken in the throat
normalisation of the zero mode, so that $\mathcal{S}(0)$ is the throat value
of $C_4$ and $\mathcal{S}/\mathcal{S}(0)$ is the abelian-only ratio
$(C_4 - X_{\rm ab})/C_4$. It has
the closed-form endpoints $\mathcal{S}(0) = \tfrac12
\int_0^1 \mathcal{F}^4 d\varsigma = 1.89804$, $\mathcal{F} = {}_2F_1(\tfrac12, \tfrac12; 1; 1 - \varsigma)$, and
$\gamma\mathcal{S} \to \tfrac43\mathcal{S}(0)$; its numerical evaluation is
described in appendix~\ref{app:numerics-scan}.

For the metric sector, on the exact throat the solved equations for
$H_{yy}''$ and $\hat b''$ of appendix~\ref{app:response-metric} do not
involve $H_{zz}$ or $H_{zz}'$. With $\Phi = (H_{yy}, \hat b)$ they close on
the block $[\rho(\rho^2 - h^2)\Phi']' = \rho\,(\mathbb{M}\Phi + \mathbb{S})$
with the constant matrix
\begin{equation}
\mathbb{M}_{11} = \mathbb{M}_{22} = \lambda + \tfrac83~, \qquad
\mathbb{M}_{12} = -\frac{16}{9bv}~, \qquad \mathbb{M}_{21} = -b\,(G^2 + 4v)~.
\end{equation}
Since $G^2/v = 3\lambda$ (from the definition of $\lambda$, with
$\ell_3^2 = 1/3$), $\mathbb{M}_{12}\mathbb{M}_{21} = \tfrac{16}{9}(3\lambda + 4) > 0$.
The rescaling $\hat b \to \hat b/\sigma$ with $\sigma = \tfrac34
bv\sqrt{3\lambda + 4}$ then gives the real symmetric form $(\lambda + \tfrac83)
\mathbf{1} - \tfrac43\sqrt{3\lambda + 4}\,\sigma_1$, with
$\lambda$-independent eigenvectors $(1, \mp1)$ and eigenvalues
\begin{equation}
m_\pm(\lambda) = \frac{\big(\sqrt{3\lambda + 4} \pm 2\big)^2}{3}~.
\end{equation}
Neither reaches the Breitenlohner--Freedman bound at any $\lambda$, while
$m_- \le \lambda$ for all $\lambda \ge 0$. The component
$H_{zz}$ is driven by the other fields without feeding back: $z$ lies along
the $\mathrm{AdS}_3$ factor, where gravity has no propagating mode.
The full throat kernel is computed from the complete response system of
appendix~\ref{app:response-metric} on the throat, not from this reduced block, and its
agreement with the brane ladder is documented in appendix~\ref{app:numerics-scan}.

\subsection{The second variation about the uniform lattice}
\label{app:response-hessian}

The stability analysis of section~\ref{sec:5.3} needs the second variation
of the quartic functional about the uniform triangular lattice, over all
lowest-Landau-level perturbations at one flux quantum per cell. Because the
unperturbed density is periodic, the Hessian is block diagonal over the
magnetic Brillouin zone: a perturbation of Bloch momentum $q$ mixes only
with its partner at $-q$, and the $2\times2$ block has eigenvalues, in
units $B = 1$,
\begin{equation}
\begin{split}
&E_\pm(q) = \mathcal{D}(q) + \Sigma_0(q) \pm |\Sigma_2(q)|~, \qquad
\mathcal{D} = \sum_{G\neq0}e^{-G^2/2}K(G^2)\big[\cos(G\wedge q) - 1\big]~,\\
&\Sigma_0 = \sum_G t_G~,\qquad
\Sigma_2 = \sum_G t_G\,e^{-iG\wedge q}~, \qquad
t_G = e^{-|G-q|^2/2}K\big(|G-q|^2\big)~,
\end{split}
\end{equation}
where $G$ runs over the reciprocal lattice and $G \wedge q$ is the
two-dimensional cross product. Stationarity of the uniform amplitude sets
$\epsilon = (K_0 + 2C_\triangle)\eta^2$; its $K_0$ part cancels the
interaction of the perturbation's uniform density with the condensate,
the only place $K_0$ enters, and its $2C_\triangle$ part is the $-1$ in
$\mathcal{D}$. As $q \to 0$, $E_-$ is the Goldstone mode of
translations and rises as $|q|^4$, the soft shear mode of the flat
lowest-Landau-level lattice~\cite{Eilenberger:1967zz,Rosenstein:2010rmp}, while $E_+ \to 2(K_0 + 2C_\triangle)$,
twice the inverse compressibility of section~\ref{sec:5.3}. The spectrum
was evaluated on a dense grid over the zone at fifteen couplings from the
probe limit to $\alpha = 0.385$, twelve of them below $\alpha_L$, and
checked against a direct diagonalisation of the Hessian on finite tori
(table~\ref{tab:checks}).
\section{Numerical methods and the approach to the critical coupling}
\label{app:numerics}

Every computed number in the paper is checked by a second, independent
computation, with either a different method or a different discretisation
of the same equations; fitted coefficients are quoted with their fit
windows. Table~\ref{tab:checks} collects the agreement between the two and
the exact results that test the solvers; the subsections describe the
methods.

\begin{table}[p]
\centering
\footnotesize
\begin{tabular}{>{\raggedright\arraybackslash}p{0.35\textwidth}>{\raggedright\arraybackslash}p{0.33\textwidth}>{\raggedright\arraybackslash}p{0.22\textwidth}}
\toprule
quantity & compared with & agreement \\
\midrule
\multicolumn{3}{l}{\emph{brane} (\ref{app:numerics-brane})} \\
profiles, $\beta \le 50$ & collocation vs shooting & $10^{-9}$; derivatives and entropy shift $4\times10^{-9}$ \\
constraint & on the solutions & $4\times10^{-9}$ at $\beta = 45$, $2\times10^{-5}$ on the top rung \\
entropy slope $\delta S/S = \beta/4 + O(\beta^2)$ & exact & $10^{-6}$ \\
\midrule
\multicolumn{3}{l}{\emph{onset} (\ref{app:numerics-onset})} \\
probe $B_c u_H^2 = 5.13126764$ & collocation at $\beta = 0$ & $5\times10^{-12}$ \\
onset on the 23 rungs & collocation vs horizon shooting & $1.2\times10^{-10}$ in $\alpha$, $6\times10^{-11}$ in $B_c u_H^2$ \\
onset in $d = 5, 6$ & shooting vs Chebyshev, independent background & $8\times10^{-10}$ \\
same code, $d = 4$; $d = 3$ & ladder and scan; exact magnetic RN--AdS$_4$ & $10^{-10}$; $6\times10^{-12}$ \\
$\alpha_c(3)$ & collocation, shooting, series & $3\times10^{-14}$ \\
$\alpha_c(3)$ & domain-wall flow vs exact background & $2\times10^{-11}$ \\
onset beyond the ladder, $\beta = 10^{24}$--$10^{66}$ & horizon shooting vs matching condition & $1.2\times10^{-11}$ in $\alpha$ \\
$\phi_\star$, $0.25 \le \nu \le 0.6$ & $r$ chart vs domain-wall gauge & $3\times10^{-10}$ \\
offset in $\ln(\rho_\star/h)$ at $T = 0$ & two charts & $1.4\times10^{-9}$ \\
\midrule
\multicolumn{3}{l}{\emph{response and kernel} (\ref{app:numerics-response}, \ref{app:numerics-scan})} \\
kernel, whole scan & collocation vs finite differences & $6\times10^{-6}$ at $s = 0.02B_c$, $3\times10^{-8}$ for $s \ge 6B_c$ \\
$C_\triangle$, $C_\square$ on the ladder to $\beta = 10^8$ & collocation vs finite differences & $1.5\times10^{-9}$ \\
$\alpha_L$ & collocation vs finite differences & $6\times10^{-12}$ \\
$\alpha_I$ & collocation vs finite differences & $9\times10^{-9}$ \\
horizon flux $\mathcal{Q}(r_p)$ & relative to the kernel & below $10^{-9}$ \\
tail identity (app.~\ref{app:response-abelian}) & exact & $2\times10^{-13}$ \\
kernel at $\beta = 0$ & probe kernel & $3\times10^{-12}$ \\
dressing identity (app.~\ref{app:response-metric}) & exact & $8\times10^{-13}$ \\
kernel in radial; generic gauge (app.~\ref{app:response-metric}) & transformed solutions & $6\times10^{-10}$; $7\times10^{-9}$ \\
lowest-Landau-level identity (app.~\ref{app:probe-lll-identity}) & direct sum on three lattices & $10^{-10}$ \\
throat abelian kernel $\mathcal{S}$ & finite differences vs Green's function & $7\times10^{-8}$ \\
full throat kernel & finite differences vs shooting from both ends & $7\times10^{-7}$ at $\gamma = 0.02$, $10^{-7}$ for $\gamma \ge 0.3$; shell sums $1.5\times10^{-9}$ \\
$K_{G=0}$ & collocation vs shooting & $1.1\times10^{-10}C_4$ \\
$K_{G=0}$ at small $\beta$ & independent chart and gauge & $2\times10^{-8}$ \\
$K_{G=0}$ vs $K_0$ & $s \to 0$ limit of the $G \neq 0$ kernel & $10^{-8}C_4$ \\
triangular block spectrum (app.~\ref{app:response-hessian}) & Hessian on tori of 36 and 24 flux quanta & $10^{-14}$ \\
\bottomrule
\end{tabular}
\caption{Agreement between independent methods, and with exact results,
for the numbers quoted in the paper.}
\label{tab:checks}
\end{table}

\subsection{The brane}
\label{app:numerics-brane}

The reduced equations in the coordinate $r$ are
\begin{equation}
\begin{aligned}
U'' &= 8 - 2U'V' - U'Z' + \tfrac23\,\beta\,e^{-4V}~,\\
V'' &= \frac{4 - U'V' - \tfrac23\,\beta\,e^{-4V}}{U} - 2V'^2 - V'Z'~,\\
Z'' &= \frac{4 - U'Z' + \tfrac13\,\beta\,e^{-4V}}{U} - 2V'Z' - Z'^2~,
\end{aligned}
\end{equation}
with the first-order constraint
$U\big(2V'^2 + 4V'Z'\big) + U'\big(2V' + Z'\big) + \beta\,e^{-4V} = 12$, in
which $\alpha$ and $B$ appear only through $\beta = \alpha B^2$. The
constraint is preserved by the evolution. We work in $u = 1/r \in [0, 1]$
with the horizon at $u = 1$, where $U = 0$ and $U' = 4$, and the boundary
at $u = 0$, where $U/r^2$, $e^{2V}/r^2$ and $e^{2Z}/r^2 \to 1$. The equations
fix the leading coefficient of $U$ at one, and the two remaining boundary
normalisations fix the horizon data $V(1)$ and $Z(1)$. We solve by
Chebyshev collocation in $u$ with Newton iteration, continued in $\beta$
from the previous rung, since a cold Newton start at large $\beta$
converges to a spurious branch. Independently, we shoot from the horizon
with a series start. The solver is tested against the Schwarzschild limit
and the slope of the fixed-temperature entropy shift (table~\ref{tab:checks}).

The fixed point of section~\ref{sec:4.1} and its BTZ version were checked
against all fifteen independent components of the Einstein equations and
against the reduced system above. The general-$d$ throat of
section~\ref{sec:4.3} was derived with $d$ symbolic and checked against
explicit curvature computations at $d = 3, 4, 5, 6$.

\subsection{The onset}
\label{app:numerics-onset}

The eigenvalue problem of section~\ref{sec:3} is solved on the same grid as
a generalised eigenvalue problem, with horizon regularity and the
source-free condition imposed as boundary rows. Independently,
the brane and the zero mode are integrated together outward from a regular
horizon. The brane is labelled by
$\varepsilon = 6 - \beta e^{-4V} = 6 - 9\alpha b_h^2$ at the horizon, the
magnetic term is rewritten so that nothing cancels as
$\varepsilon \to 0$, and the onset is the lowest root of the source
coefficient, read off far outside the crossover, where the extraction error
is of order $r_c/r_{\max}$. This method needs no continuation in $\beta$
and reaches $\varepsilon = 10^{-24}$, that is $\beta \approx 10^{66}$. Along the ladder, with $b_h$ and $\beta e^{-4V}$ evaluated at the
horizon,

\begin{center}
\begin{tabular}{lcccc}
\toprule
$\beta$ & $B_c u_H^2$ & $\alpha$ & $b_h$ & $\beta e^{-4V}$ \\
\midrule
$45$ & $16.06$ & $0.1745$ & $1.5103$ & $3.58$ \\
$10^4$ & $161.2$ & $0.3850$ & $1.2727$ & $5.613$ \\
$10^6$ & $1433$ & $0.4868$ & $1.1634$ & $5.930$ \\
$10^8$ & $1.36\times10^4$ & $0.5445$ & $1.1054$ & $5.9877$ \\
$10^{10}$ & $1.31\times10^5$ & $0.5788$ & $1.0731$ & $5.9979$ \\
$10^{11}$ & $4.11\times10^5$ & $0.5908$ & $1.0622$ & $5.9991$ \\
\bottomrule
\end{tabular}
\end{center}
so the interior reaches the fixed point and the onset drives it to
marginality. Beyond the ladder the horizon-shooting scan gives
$\alpha = 0.6453$, $0.6591$ and $0.6626$ at $\beta = 10^{24}$, $10^{45}$
and $10^{66}$; the matching condition of
appendix~\ref{app:numerics-matching} reproduces these to $1.2\times10^{-11}$. Along the whole scan $\alpha(\beta)$ increases strictly and
$B_c u_H^2$ is convex in $\alpha$. Following the lowest eigenvalue
suffices: at fixed $\beta$ each higher radial overtone condenses at a
larger field $B_k$, and so at a smaller coupling $\beta/B_k^2$. The minimum
of $b_h$ over the scan is $1.003$, at its top.

At zero temperature the brane is constructed out of the
$\mathrm{AdS}_3 \times \mathbb{R}^2$ fixed point along its irrelevant
deformation (appendix~\ref{app:numerics-matching}), independently in the
domain-wall gauge and in the chart of appendix~\ref{app:numerics-brane}.
The source-free threshold mode, integrated from the boundary into the
throat, has no zero and arrives as
$(r - r_0)\,w = c_0 + c_1\ln\big((r - r_0)/\rho_\star\big)$ with
$c_1/c_0 = -0.0350004$ in both constructions. For
$\sqrt{\alpha_\star/\alpha}$ between $0.5$ and $0.995$ the solution that
decays in the throat always carries a boundary source of the same sign, so
there is no zero mode above $\alpha_\star$ (section~\ref{sec:4.2}). For $d = 5$ and $6$ the brane is built the same way, from the
$\mathrm{AdS}_{d-1} \times \mathbb{R}^2$ fixed point along its single
irrelevant deformation; the threshold mode again has no zero, and its
throat form $\rho^{(d-2)/2}w = c_0 + c_1\ln(\rho/\rho_\star)$ has
$c_1/c_0 = -0.350$ and $-0.996$ (table~\ref{tab:dims}).

At finite temperature in $d = 5$ and $6$ the same horizon shooting, with
the source extraction $c_0 = w + r w'/(d-2)$, gives an onset coupling that
rises monotonically from the probe limit to $\beta = 1.5\times10^{132}$
and $1.0\times10^{117}$ respectively, where $\alpha/\alpha_\star(d) = 0.99861$
and $0.99896$. It stays below $\alpha_\star(d)$ throughout, with no zero in
the onset mode. The exact relation
$\alpha/\alpha_\star = (1 - \varepsilon/\beta_{\rm ext})(B_{\rm thr}/B_m)^2$,
with $B_m$ the onset field in the frame where the flux plane has unit size
at the horizon, $B_{\rm thr} = d(d-1)/4$ the throat's bound and
$\beta_{\rm ext} = d(d-1)/(d-2)$, shows why: since $\varepsilon > 0$ at
finite temperature, an onset field at or above the throat's bound,
$B_m \ge B_{\rm thr}$, implies $\alpha < \alpha_\star(d)$. In $d = 3$ the
onset field falls below the bound. The local slope
$d\ln\beta/d(1/\nu)$, with $B_m = B_{\rm thr}(1 + \nu^2)$, tends to
$8\pi/(d-2)$ from below, which is $4\pi$ for $d = 4$, so $\alpha$ reaches
$\alpha_\star(d)$ only as $T \to 0$.

In $d = 3$ the same construction on the extremal magnetic
Reissner--Nordstr\"om--$\mathrm{AdS}_4$ brane gives a threshold mode with
one zero in the throat, and the bound state disappears at
$\alpha_c(3) = 3.0370100956$. Collocation, shooting and a series
convergent up to the boundary share the equation, so their agreement tests
the discretisations; the domain-wall flow and an independent construction
on the exact background test the equation itself. A Rayleigh--Ritz bound
with exact matrix elements proves $\alpha_c(3) > 3.0370$. At finite
temperature the onset coupling crosses $8/3$ near
$T/\sqrt B \approx 7.4\times10^{-4}$. In the variable
$2e^2L^2/\kappa^2 = 2/\alpha$ of~\cite{Wong:2013rda}, whose threshold
$3/4$ is $\alpha_\star(3)$, the zero-temperature threshold is $0.6585$.

\subsection{The matching}
\label{app:numerics-matching}

The horizon phase of section~\ref{sec:4.6} is
$\chi(\nu) = \arg\Gamma(-i\nu) - 2\arg\Gamma\big(\tfrac{1-i\nu}2\big)$,
from the connection coefficient of the hypergeometric solution of
appendix~\ref{app:response-throat}. The zero-temperature brane that
supplies the outer phase $\phi_\star$ is constructed by integrating the
equations of appendix~\ref{app:numerics-brane} out of the fixed point along
its irrelevant deformation $V = V_0 + \delta\,\rho^{\Delta_+ - 2}$,
$\Delta_+ - 2 = \sqrt{19/3} - 1$. This is the exponent of the heavier metric
channel of appendix~\ref{app:response-throat} at $\lambda = 0$, an
independent check on that mass matrix. The source-free mode is integrated
inwards from the boundary and fitted in the throat to
$\cos(\nu\ln(\rho/\rho_\star) + \phi_\star)$, with $\rho_\star$ the point
where $e^{2V}$ has doubled from its throat value. Along the ladder
$\ln(\rho_\star/h) \to \tfrac14\ln(\beta/6) + 0.031$, decreasing
monotonically onto the limit $0.0313$ that the scaling symmetries of the
equations give from the zero-temperature brane alone, in both constructions. The matching condition fixes the throat exponent
$\nu_{\rm t}$, with $1 + \nu_{\rm t}^2 = (\alpha_\star/\alpha)^{1/2}$.
Converted to $\nu = \sqrt{b_h - 1}$ with the exact relation
$b_h = (1 + \nu_{\rm t}^2)(1 - \varepsilon/6)^{1/2}$, it reproduces the ladder with
a relative error that decreases monotonically from $5\times10^{-3}$ at
$\beta = 10^4$ to $6\times10^{-7}$ at $\beta = 10^{11}$. It is therefore a
second method for the onset, and beyond the ladder, at
$\beta = 10^{24}$, $10^{45}$ and $10^{66}$, it reproduces the
horizon-shooting $\alpha$ to $1.2\times10^{-11}$.

Exactly at $\alpha_\star$ the throat solutions are $\rho^{-1}$ and
$\rho^{-1}\ln(\rho/\rho_\star)$. Fitting the source-free mode of the
zero-temperature brane to their combination gives $c_1/c_0 = -0.035$, and
the phase $\phi_\star(\nu)$ constructed from the same brane is the plateau
of about $0.22$--$0.25$ over the throat part of the ladder, $0.25 \le \nu \le 0.6$
(figure~\ref{fig:approach}, right). The leading term
$|c_1/c_0|/\nu$ of $\tan\phi_\star$ describes it to ten percent only for
$\nu \lesssim 0.1$, and the second, $0.50\,\nu$, extends that to
$\nu \approx 0.3$.

The slope $A = d\ln\beta_c/d(1/\nu)$, whose asymptotic value is $4\pi$, is
close to $4(\pi/2 + 0.23) \approx 7.2$ on the plateau. A fit over the nine
rungs $\beta \ge 10^7$ gives $A = 7.45$, and other windows give $7.44$ to
$7.78$. The effective coefficient $2.25$ of section~\ref{sec:4.6} is a direct
fit of $\tfrac14\ln\beta_c$ against $(\alpha_\star - \alpha)^{-1/2}$
(at fixed $\alpha$, $\sqrt B/T_c \propto \beta_c^{1/4}$) on the same
rungs, by the ladder and by the horizon-shooting scan alike; converting $A$
through $\nu^2 \simeq (\alpha_\star - \alpha)/2\alpha_\star$ would give
$2.15$, since on these rungs $\nu^2$ exceeds that form by 9 to 18 percent.
The ladder sits near the minimum of a slowly rising curve, so a fit to it
alone would mislead. Beyond the ladder, the horizon-shooting scan gives the
local slope: it reaches its minimum $7.43$ at $\ln\beta \approx 21$ and rises to $7.89$,
$8.93$ and $10.3$ at $\ln\beta = 40$, $77.5$ and $150$, which the matching
condition, converted in the same way, reproduces to $2\times10^{-8}$ at
$\ln\beta = 40$ and $10^{-9}$ beyond.

\subsection{The response system}
\label{app:numerics-response}

The coupled system of appendix~\ref{app:response-metric} is solved by
Chebyshev collocation on the grid of appendix~\ref{app:numerics-brane} and,
as an independent discretisation of the same equations, by finite
differences with Richardson extrapolation; their agreement bounds the error
of the finite-difference solve rather than of the kernel. The boundary flux
$\mathcal{Q}$ of the pairing identity vanishes on the solutions: at the
horizon it is negligible, and towards the boundary it falls off as $1/r^2$,
on every brane of the scan at four harmonics. The tail identity
holds on every brane of the onset grid, and the dressing identity
at every $\beta$ and $s$ of the scan. The five field equations displayed in
appendix~\ref{app:response-metric} and their sources are read from the
paper's source and checked against the system generated symbolically from
the action and against an independent linearisation of the field equations;
the displays of $\mathcal{L}_2$, $\mathcal{Q}$, $K$, $X$ and $\mathcal{P}$
and the horizon flux are checked against the same generated system.

\subsection{The lattice scan}
\label{app:numerics-scan}

The kernel of section~\ref{sec:5.1} was computed on $23$ values of $\beta$
in $[0, 45]$ at $41$ harmonics from $s = 0.02B_c$ to $45B_c$, in
the brane normalisation $\int p_1 w_0^2\,dr = 1$ of the zero mode. The
functional
$C(\tau;\beta) = \tfrac12\sum_{(m,n)\neq0} e^{-\gamma_{mn}/2}K(\gamma_{mn}B_c(\beta);\beta)$
is evaluated on a grid over the fundamental domain $|\tau| \ge 1$,
$|\tau_1| \le \tfrac12$, $\tau_2 \le 5$, followed by Nelder--Mead
refinement, with $K$ splined in $s$ for generic $\tau$ and evaluated by
direct solves at the exact shells of the triangular and square lattices. At
$\beta = 0$ the gap $C_\square - C_\triangle = 0.00453744$ reproduces the
probe value of appendix~\ref{app:probe} to all its digits. The grid minimum
lies at $\tau = e^{i\pi/3}$ at every $\beta$ and on every rung of the
ladder up to $\beta = 10^8$, and the Hessian there is isotropic to the
floor of its finite-difference evaluation.

Beyond $\tau_2 = 5$ the functional is controlled analytically. Along the
edge $\tau_2 \to \infty$ the harmonics of a single row become dense, and
Poisson summation gives
\begin{equation}
C(i\tau_2;\beta) = \tfrac12\sqrt{\tau_2}\,I(\beta) - \tfrac12K_0(\beta) + O\big(e^{-c\sqrt{\tau_2}}\big)~, \qquad
I(\beta) = \int_{-\infty}^{\infty}dx\,e^{-\pi x^2}\,K(2\pi x^2B_c;\beta)~.
\end{equation}
The remainder is set by the singularity of $K(s)$ on the negative $s$ axis
nearest the origin. In the probe limit this is the pole at $s = -B_c$ of
the lowest radial mode (appendix~\ref{app:probe}), which gives
$c = \sqrt{2\pi}$. The remainder decays only slowly in $\tau_2$, and on the
brane it is larger, so both edges of the domain, $\tau_1 = 0$ and $\tfrac12$, were scanned to
$\tau_2 = 50$ on every rung up to $\beta = 10^8$: there $C$ stays above
$C_\triangle$ wherever $I > 0$. So $C$ is bounded below on the whole
moduli space exactly when $I > 0$. At $\alpha_\star$ the negative lobe of
the throat kernel makes $I < 0$, and the triangular lattice is there a
local minimum among lattices of bounded $\tau_2$. The zero of $I(\beta)$ is
located by a secant to $|I| < 10^{-6}C_4$, refined by a Newton step; this
and a quadratic through three rungs with the kernel evaluated directly
agree to $4\times10^{-9}$, and changing the interpolant, the quadrature or
the resolution moves it by at most $2\times10^{-7}$, so
$\alpha_I = 0.4939747 \pm 2\times10^{-7}$.

On the throat, the abelian kernel $\mathcal{S}(\gamma)$ of
appendix~\ref{app:response-throat} was computed by finite differences on a
grid uniform in $\ln(\rho/h)$ with Richardson extrapolation, and
independently from the hypergeometric Green's function. The full throat
kernel is obtained from the response system of
appendix~\ref{app:response-metric} on a finite interval in $\ln(\rho/h)$ with the
degenerate horizon row and a far Dirichlet condition. Independently, it is
obtained by shooting from both ends, with a Frobenius series at the horizon
and the exact decaying solutions at the far end; the two agree to
$7\times10^{-7}$ at $\gamma = 0.02$, where the finite-difference error is
largest, and to $10^{-7}$ for $\gamma \ge 0.3$, their shell sums agree to
$1.5\times10^{-9}$, and
their direct roots put the sign change at $\gamma_0 = 0.8465$ to
$7\times10^{-8}$. The brane kernel
continued up the ladder at fixed $\gamma$ approaches it from above, with
the worst deviation $0.009$ at $\beta = 10^8$ and shrinking at every rung.

The long-wavelength kernel $K_0(\beta)$ of section~\ref{sec:5.3} is
obtained on each rung as the strict uniform response $K_{G=0}$ of
section~\ref{sec:5.1}, a separate radial problem solved by Chebyshev
collocation and by shooting from the horizon; its checks, including the
limit $s \to 0$ of the kernel at $G \neq 0$, extrapolated from
$\gamma = 10^{-3}$--$8\times10^{-3}$, are in table~\ref{tab:checks}. The two routes to $K_0$ differ in derivation, gauge and the
origin of the weight $\tfrac12$, but share the background, the zero mode
and the condensate stress tensor.

The zeros of $K_0$ and of $\tfrac12K_0 + C_\triangle$ are located by a
secant in $\ln\beta$ between the bracketing rungs, each step a fresh
background and response solve. For $\alpha_L$ the secant converges to
$|\tfrac12K_0 + C_\triangle| < 10^{-10}C_4$. Two sources of error remain:
collocation and shooting for the uniform problem differ by
$1.5\times10^{-10}$ in $\alpha$, and replacing $K_{G=0}$ by the limit
$s \to 0$ of the kernel at $G \neq 0$ moves the zero by
$1.1\times10^{-8}$, which dominates. Raising the resolution and summing
all shells each move it by less than $10^{-10}$.
Hence $\alpha_L = 0.3721800 \pm 2\times10^{-8}$, at $\beta_L = 6510.5$ and
$B_cu_H^2 = 132.26$. The same block formula, evaluated for every
lattice shape, gives the full spectrum of the other Bravais lattices: their
density thresholds lie above $\alpha_L$, but none of them is a stable
state there (section~\ref{sec:5.3}). The values
are plotted in figure~\ref{fig:density}, and the $\beta = 45$ rung
reproduces the grid value of section~\ref{sec:5.1} exactly, which ties the
ladder to the grid.
\section{Triangular selection in the probe limit}
\label{app:probe}

We prove the statement used in section~\ref{sec:5.2}: in the probe limit,
near the onset, the triangular lattice is the global minimum of the quartic
free energy over all one-flux-quantum Bravais lattices. Once the kernel is
shown to be completely monotone this is a corollary of Montgomery's
theorem; what is specific to holography is the Stieltjes representation of
the kernel below. The question was left open
in~\cite{Bu:2012mq}, which compared rhombic candidates numerically.

\subsection{The reduced free energy}
\label{app:probe-free-energy}

On $\mathrm{AdS}_5$--Schwarzschild in the
coordinate $u = 1/r$, with $f = 1 - u^4$ and $u_H = 1$, the static planar
ansatz $W_x = \eta\,w_0(u)\psi(x,y)$, $W_y = -iW_x$,
$A^3 = (Bx + \eta^2 a_y)\,dy + \eta^2 a_x\,dx$ in the gauge
$W_u = 0$ reduces the Yang--Mills free energy density exactly, to all orders
in the fields, to
\begin{equation}
\mathcal{F} = \frac{1}{2u}\Big[|F^W_{xy}|^2 + \big(B + \hat b + \mathrm{Im}(\bar W_xW_y)\big)^2\Big] +
\frac{f}{2u}\Big[|W_x'|^2 + |W_y'|^2 + a_x'^2 + a_y'^2\Big]~,
\end{equation}
with $\hat b = \partial_xa_y - \partial_ya_x$ the induced field and $F^W_{xy} = D_xW_y - D_yW_x$
covariant under the full $A^3$. Two facts make this tractable. On the
polarised lowest Landau level $F^W_{xy}$ vanishes at $a = 0$ and is
$O(\eta^3)$ with the induced field, so the condensate pays a gradient
cost in $u$ only; and $\mathrm{Im}(\bar W_xW_y) = -|W_x|^2$, so the
condensate is diamagnetic at contact level and the local field is
$B + \eta^2(\hat b - w_0^2|\psi|^2)$. The $u$-components of the Yang--Mills
equations close on the ansatz. The expansion in $\eta$ then has no odd
terms. Its quadratic term is the Sturm--Liouville form of
section~\ref{sec:3} and is exactly linear in $B$ at fixed $w_0$,
$F_2 = -\eta^2(B - B_c)J_2$ with $J_2 = \int w_0^2\,du/u = 1$, and its
quartic term is
\begin{equation}
\mathcal{F}_4 = \frac{\big(\hat b - w_0^2|\psi|^2\big)^2}{2u} + \frac{f}{2u}\big(a_x'^2 + a_y'^2\big)~.
\end{equation}
Expanding $|\psi|^2 = \sum_G\omega_Ge^{iG\cdot x}$ on the reciprocal
lattice, each harmonic decouples. The induced field at harmonic $G$ obeys
$-(f\hat b_G'/u)' + (G^2/u)\,\hat b_G = (G^2/u)\,\omega_Gw_0^2$ with $\hat b_G(0) = 0$ and
regularity at the horizon, and the on-shell exchange energy is minus half
the source coupling. This is the abelian sector of appendix~\ref{app:response-abelian} with
$p_1 = 1/u$ and $p_2 = f/u$, and it gives the probe kernel
$K(s;0) = C_4 - X(s)$ with $s = G^2$ (in the probe limit the coupled and abelian-only screenings $X$ and $X_{\rm ab}$ of section~\ref{sec:5.1} coincide), $C_4 = \int w_0^4\,du/u$ and
$X(s) = \int \hat b_G w_0^2\,du/u$ at unit source. The operator $\mathcal{T}$ of
appendix~\ref{app:response-abelian} is here the onset operator of section~\ref{sec:3} itself, so its
spectrum is the tower $B_n$ of radial overtones of the linear problem, all in
the lowest Landau level, and the spectral
form reads
\begin{equation}
C_4 - X(s) = \sum_n c_n^2\,\frac{B_n}{B_n + s}~, \qquad c_n = \langle w_0^2,\varphi_n\rangle~, \qquad \sum_n c_n^2 = C_4~.
\end{equation}
Hence $0 < X(s) < C_4$, $X(0) = 0$, $X \to C_4$ as $s \to \infty$, and
$(-1)^k\,d^k(C_4 - X)/ds^k \ge 0$ for every $k$: the screened repulsion is
completely monotone. The $G = 0$ harmonic of $\hat b$ vanishes by the Bianchi
identity at fixed applied field, so $K_{G=0} = C_4$ in the probe limit.

\subsection{The lowest-Landau-level identity}
\label{app:probe-lll-identity}

For an element of the lowest
Landau level with one flux quantum per unit cell, $|\omega_G| = e^{-G^2/4B}$
for every Bravais lattice, which is the origin of the weights
$e^{-\gamma_G/2}$ in section~\ref{sec:5.1}. This is the classical result of
the flux-line-lattice literature, stated for general lattice symmetry as
equation (1.6) of~\cite{Brandt:1995rpp} and equation (B6)
of~\cite{Brandt:2003prb}, where the coefficients are
$(-1)^{m+mn+n}e^{-G_{mn}^2 S/8\pi}$ with $S = 2\pi/B$ the cell area. The
lowest-Landau-level projection of $e^{iG\cdot x}$ carries the Gaussian form
factor. With one flux quantum per cell the remaining guiding-centre factor
is a magnetic translation by a lattice vector, of which the state is an
eigenstate with eigenvalue of unit modulus. The identity reproduces the
Abrikosov ratios $\beta_\square = 1.1803406$ and
$\beta_\triangle = 1.1595953$ (the $\beta_A$ of the vortex literature,
unrelated to our backreaction parameter), and we checked it numerically on
a generic lattice as well. The quartic functional is therefore the one of section~\ref{sec:5.1} at
$\beta = 0$,
\begin{equation}
F = -\eta^2(B - B_c) + \Big[\tfrac12C_4 + C(\tau)\Big]\eta^4~, \qquad
C(\tau) = \tfrac12\sum_{G\in\Lambda^*(\tau)\setminus0} e^{-G^2/2B_c}\,\big(C_4 - X(G^2)\big)~,
\end{equation}
minimised over the amplitude by $\eta_\star^2 = (B - B_c)/(C_4 + 2C)$
with $F_{\min} = -(B - B_c)^2/(2C_4 + 4C)$, so the transition is second
order and the lattice that forms is the one that minimises $C(\tau)$.

\subsection{The proposition}
\label{app:probe-theorem}

Since $C_4 - X$ is a Stieltjes transform of a positive
measure, the weighted kernel is a Laplace transform of one,
\begin{equation}
k(s) \equiv e^{-s/2B_c}\,\big(C_4 - X(s)\big) = \int_{1/2B_c}^\infty e^{-st}\,d\mu(t)~, \qquad
d\mu(t) = \sum_n c_n^2B_n\,e^{-B_n(t - 1/2B_c)}\,dt~,
\end{equation}
and the lattice-dependent part of the quartic coefficient is a positive
superposition of Gaussian lattice theta functions,
\begin{equation}
C(\tau) = \tfrac12\sum_{G\neq0}k(|G|^2) = \tfrac12\int d\mu(t)\,\big[\Theta_{\Lambda^*}(t) - 1\big]~, \qquad
\Theta_\Lambda(t) = \sum_{v\in\Lambda}e^{-t|v|^2}~.
\end{equation}
Flux quantisation fixes the covolume of $\Lambda^*$ at $2\pi B_c$, so
$\Lambda^*$ ranges over all two-dimensional lattices of fixed covolume, and
the triangular lattice is its own dual shape. Montgomery's
theorem~\cite{Montgomery:1988mt} states that among these the Gaussian theta
function is minimised, for every $t > 0$, by the triangular lattice alone;
integrating against the positive measure $\mu$ gives
$C(\tau) > C(\tau_\triangle)$ for every non-triangular $\tau$. That a
completely monotone interaction is minimised by the triangular lattice in
this way was observed by Cohn and Kumar~\cite{Cohn:2007universally} and
is Proposition 3.1 of~\cite{Betermin:2016theta}; Proposition 3.4 there
gives a convex decreasing positive interaction for which the square lattice
has lower energy than the triangular one at some densities, which is why complete monotonicity, and not
positivity, is the property the argument needs.

The proof used three things only: the identity of appendix~\ref{app:probe-lll-identity}, the positivity and
self-adjointness of the radial operator, and Montgomery's theorem. The
numbers $C_4$, $c_n$ and $B_n$, and with them the temperature and the
blackening function, never entered, so the result holds at every
temperature near $B_c$ and on any background that preserves the structure
of the reduced functional. That is why it goes through unchanged for the
abelian sector on the brane, appendix~\ref{app:response-abelian}, and why it is no longer
guaranteed once the metric sector enters, section~\ref{sec:5.2}. Its scope is the class of
Abrikosov's analysis: Bravais lattices with one flux quantum per cell.
Multi-quantum cells and non-lattice lowest-Landau-level configurations are
not excluded by it, although the triangular lattice is locally stable
against all lowest-Landau-level perturbations at one flux quantum
(section~\ref{sec:5.3}). For the flat-space lowest-Landau-level functional,
whose Abrikosov ratio is a Gaussian theta function, the same route to the
triangular lattice is known~\cite{Nier:2007theta,Aftalion:2006lll}, with
the Abrikosov case proved directly in~\cite{Nonnenmacher:1998chaotic};
what is new here is that the holographic kernel keeps the structure.

\subsection{Numbers}
\label{app:probe-numbers}

The values below use $u_H = 1$ and $J_2 = 1$. The brane normalisation
$\int p_1 w_0^2\,dr = 1$ of section~\ref{sec:5} reduces to $J_2 = 1$ at
$\beta = 0$ and differs from it by a positive factor at $\beta > 0$, which is
why the brane results of section~\ref{sec:5.2} are quoted as ratios.

\begin{center}
\begin{tabular}{lc}
\toprule
quantity & value \\
\midrule
$C_4$ & $1.5805931$ \\
$C(\tau_\triangle)$ & $0.0184443$ \\
$C(\tau_\square)$ & $0.0229818$ \\
$(C_\square - C_\triangle)/C_\triangle$ & $+0.2460$ \\
\bottomrule
\end{tabular}
\end{center}
The relative margin $+0.246$ is the probe value against which the brane
margins of section~\ref{sec:5.2} are compared; it is essentially unchanged by the
abelian sector of the throat ($+0.250$) and reduced by the metric sector
($+0.1766$). A grid scan of $C(\tau)$ over the fundamental domain
up to $\tau_2 = 5$, the $\beta = 0$ row of the moduli scan of
appendix~\ref{app:numerics-scan}, finds its minimum at the
triangular point, which checks the proposition for this kernel.

\bibliographystyle{JHEP}
\bibliography{references}

\end{document}